\documentclass{aa}
\usepackage{txfonts}
\usepackage{graphicx}
\usepackage{natbib}
\begin{document}

\title{Signals embedded in the radial velocity noise}
\subtitle{Periodic variations in the $\tau$ Ceti velocities}


\author{M. Tuomi\thanks{\email{mikko.tuomi@utu.fi; m.tuomi@herts.ac.uk}}\inst{1,2} \and H. R. A. Jones\inst{1} \and J. S. Jenkins\inst{3} \and C. G. Tinney\inst{4,5} \and R. P. Butler\inst{6} \and S. S. Vogt\inst{7} \and J. R. Barnes\inst{1} \and \\ R. A. Wittenmyer\inst{4,5} \and S. O'Toole\inst{7} \and J. Horner\inst{4,5} \and J. Bailey\inst{4} \and B. D. Carter\inst{3} \and D. J. Wright\inst{4,5} \and G. S. Salter\inst{4,5} \and D. Pinfield\inst{1}}

\institute{
University of Hertfordshire, Centre for Astrophysics Research, Science and Technology Research Institute, College Lane, AL10 9AB, Hatfield, UK
\and University of Turku, Tuorla Observatory, Department of Physics and Astronomy, V\"ais\"al\"antie 20, FI-21500, Piikki\"o, Finland
\and Departamento de Astronom\'ia, Universidad de Chile, Camino del Observatorio 1515, Las Condes, Santiago, Chile
\and School of Physics, University of New South Wales, 2052, Sydney, Australia
\and Australian Centre for Astrobiology, University of New South Wales, 2052, Sydney, Australia
\and Department of Terrestrial Magnetism, Carnegie Institute of Washington, Washington, DC 20015, USA
\and UCO/Lick Observatory, Department of Astronomy and Astrophysics, University of California at Santa Cruz, Santa Cruz, CA 95064, USA
}


\date{Received XX.XX.2012 / Accepted XX.XX.XXXX}

\abstract{The abilities of radial velocity exoplanet surveys to detect the lowest-mass extra-solar planets are currently limited by a combination of instrument precision, lack of data, and ``jitter''. Jitter is a general term for any unknown features in the noise, and reflects a lack of detailed knowledge of stellar physics (asteroseismology, starspots, magnetic cycles, granulation, and other stellar surface phenomena), as well as the possible underestimation of instrument noise.}
{We study an extensive set of radial velocities for the star HD 10700 ($\tau$ Ceti) to determine the properties of the jitter arising from stellar surface inhomogeneities, activity, and telescope-instrument systems, and perform a comprehensive search for planetary signals in the radial velocities.}
{We perform Bayesian comparisons of statistical models describing the radial velocity data to quantify the number of significant signals and the magnitude and properties of the excess noise in the data. We reach our goal by adding artificial signals to the ``flat'' radial velocity data of HD 10700 and by seeing which one of our statistical noise models receives the greatest posterior probabilities while still being able to extract the artificial signals correctly from the data. We utilise various noise components to assess properties of the noise in the data and analyse the HARPS, AAPS, and HIRES data for HD 10700 to quantify these properties and search for previously unknown low-amplitude Keplerian signals.}
{According to our analyses, moving average components with an exponential decay with a timescale from a few hours to few days, and Gaussian white noise explains the jitter the best for all three data sets. Fitting the corresponding noise parameters results in significant improvements of the statistical models and enables the detection of very weak signals with amplitudes below 1 ms$^{-1}$ level in our numerical experiments. We detect significant periodicities that have no activity-induced counterparts in the combined radial velocities. Three of these signals can be seen in the HARPS data alone, and a further two can be inferred by utilising the AAPS and Keck data. These periodicities could be interpreted as corresponding to planets on dynamically stable close-circular orbits with periods of 13.9, 35.4, 94, 168, and 640 days and minimum masses of 2.0, 3.1, 3.6, 4.3, and 6.6 M$_{\oplus}$, respectively.}
{}

\keywords{Methods: Statistical, Numerical -- Techniques: Radial Velocities -- Stars: Individual: HD 10700}


\maketitle


\section{Introduction}

The improving instrumental precision and the rapidly increasing number of measurements of radial velocity (RV) target stars of different surveys are enabling the discoveries of smaller planets on longer orbits in increasing numbers \citep[e.g.][]{lovis2011,mayor2009,mayor2011}. Currently, however, the stellar noise, sometimes referred to as stellar jitter \citep[e.g.][]{wright2005}, presents the greatest obstacle in reaching the ultimate goal of being able to detect Earth-mass planets in stellar habitable zones \citep{barnes2012}. While the magnitude of this jitter is still very uncertain for all the target stars, its other properties, such as shape of the noise distribution, its variability as a function of time, and dependence on the stellar properties have not received much attention.

When analysing RV data, the common strategy is to bin the measurements made within, say, intervals of an hour, and use the resulting binned data as a basis of further statistical analyses. The motivation for this operation is the possibility of reducing the amount of short-term noise in the data \citep[e.g.][]{otoole2009}. While this approach certainly mitigates against asteroseismological signals, it results in a loss of information about the properties of stellar noise and may also prevent the detection of the faintest signals in the data because it corresponds to artificially decreasing the number of measurements. In general, RV signals are published in the literature as received from binned data but with no information about the binning techniques used, i.e. binning time-scale and uncertainty estimation. To circumvent this approach and its possible shortcomings, we study the properties of the jitter by comparing different noise models: different autoregressive (AR) and moving average (MA) models, their combinations (ARMA), and different additive white noise models.

Because it is not possible to generate artificial RV data with realistic noise properties (for the obvious reason that these properties are not known), we study a quiet star with a large amount of data and no published signals. We analyse the RVs of the nearby Solar-type star HD 10700 ($\tau$ Ceti). It is known to be a very inactive and quiescent star and its long high-precision RV data set from the High Accuracy Radial Velocity Planet Searcher (HARPS) spectrograph \citep{mayor2003} has not been reported to contain planetary signatures despite more than 4000 spectral observations \citep{pepe2011}. Also, there are two other large RV data sets of  HD 10700 available from the High Resolution Echelle Spectrograph \citep[HIRES;][]{vogt1994} on the Keck telescope and from the U.C.L Echelle Spectrograph at the Anglo Australian Telescope (AAT). We refer to the RV data from the AAT as Anglo-Australian Planet Search (AAPS) data. To verify the trustworthiness of our noise models in extracting weak signals from the data, we first test their performance by adding artificial signals to the HARPS data set for HD 10700. The models that enable the recovery of the artificial signals are then compared using the Bayesian model selection techniques to find the most accurate descriptions of these HARPS RVs. Finally, we search for periodic signatures of planetary companions in the HARPS velocities.

We also find the best noise models for the AAPS and HIRES RVs and use them in combination with the HARPS velocities in further analyses because signals that are not detected in any of these data sets might be available from the combined set because of its greater size and better phase-coverage relative to the individual data sets.

We start by describing the properties of the target star HD 10700 and the RV data in Section \ref{sec:properties} and the statistical tools (methods and models) in Section \ref{sec:modelling}. Model selection methods and our criteria for detecting signals in the RV data are presented in Sections \ref{sec:noise_selection} and \ref{sec:signal_detection}, respectively. After this, we proceed by determining which noise model performs the best in extracting artificial signals introduced into the data (Section \ref{sec:artificial}). The best noise models are then used to analyse the HARPS data (Section \ref{sec:harps_analysis}). To make sure that the signals we detect do not correspond to any activity-related phenomena, we analyse the timeseries of HARPS activity indices in Section \ref{sec:activity}. We also perform analyses of the AAPS and HIRES data sets (Section \ref{sec:other_analyses}) and the combined data (Section \ref{sec:combined_analyses}). Finally, we assess briefly the dynamical stability of the system of planet candidates we find in the combined data (Section \ref{sec:dynamics}) and present the conclusions and discussion in Section \ref{sec:discussion}.

\section{HD 10700 and radial velocity measurements}\label{sec:properties}

\subsection{Stellar properties}

HD 10700 is a very nearby Solar-type \citep[G8.5 V;][]{gray2006} star with a large Hipparcos parallax of 273.96$\pm$0.17 mas \citep{vanleeuwen2007} implying a distance of only 3.650$\pm$0.002 pc. Because only 18 stellar systems (single, double, or triple stars) are located closer to the Sun, HD 10700 is in the immediate neighbourhood of our own system. While it can readily be called a Sun-like star, HD 10700 is somewhat lighter with a mass of $0.783 \pm 0.012$ M$_{\odot}$ \citep{teixeira2009} and less luminous and also has a sub-Solar metallicity of -0.55$\pm$0.05 \citep[][ and references therein]{pavlenko2012}, which could potentially make it an ideal target for searches for low-mass planets due to the relatively higher frequency of low-mass planets around low-metallicity stars \citep{jenkins2012}.

\begin{table}
\center
\caption{Estimated stellar properties of HD 10700.\label{properties}}
\begin{tabular}{lrrrrrr}
\hline \hline
Parameter & Value & Reference \\
\hline
Spectral Type & G8.5 V & (i) \\
$\log R'_{\rm HK}$ & -4.995 & (h) \\
$\pi$ [mas] & 273.96$\pm$0.17 & (a) \\
L$_{\rm star}$ [L$_\odot$]& 0.488$\pm$0.010 & (b) \\
R$_{\rm star}$ [R$_\odot$] & 0.793$\pm$0.004 & (b) \\
M$_{\rm star}$ [M$_\odot$] & 0.783$\pm$0.012 & (b) \\
T$_{\rm eff}$ [K] & 5344$\pm$50 & (f) \\
$[$Fe/H$]$ & -0.55$\pm$0.05 & (j) \\
Age [Gyr] & 5.8 & (d) \\
$v$ sin $i$ [kms$^{-1}$] & 0.90 & (g) \\
P$_{\rm rot}$ [days] & 34 & (e) \\
\hline \hline
\end{tabular}
\tablefoot{Data from: (a) \citet{vanleeuwen2007}; (b) \citet{teixeira2009}; (c) \citet{soubiran1998}; (d) \citet{mamajek2008}; (e) \citet{baliunas1996}; (f) \citet{santos2004}; (g) \citet{santos2002}; (h) \citet{pepe2011}; (i) \citet{gray2006}; (j) \citet{pavlenko2012}.}
\end{table}

Despite the fact that HD 10700 is a target star of several RV searches for planets around nearby stars (e.g. the HARPS, Keck-HIRES, and AAPS, described in the next section), no planetary or other sub-stellar companions have been reported orbiting it \citep{wittenmyer2006,pepe2011}. However, the star has a bright debris disk, i.e. a circumstellar disk estimated to be an order of magnitude greater than the mass of Edgeworth-Kuiper belt in the Solar System, extending out to $\sim$ 55 AU \citep{greaves2004} and has been observed to have warm dust orbiting it \citep{difolco2007}. These findings promote the possibility of HD 10700 having some of this circumstellar matter in the form of planets orbiting the star as well. Combining these arguments, and noting that HD 10700 is a very inactive star \citep[a ``flat activity'' star;][]{judge2004} with $\log R_{\rm HK} = -4.955$ \citep{pepe2011}, it seems likely that either HD 10700 is pole-on \citep[as also suggested by ][though there is currently no evidence in favour of this hypothesis]{gray1994}, which, given co-planarity, prevents the detections of planets orbiting HD 10700 using the RV method (and indeed using the transit photometry), or its companions are very small and do not induce observable periodic Doppler variations to the stellar spectra. The obvious third option is that there simply are no planets orbiting HD 10700. While a plausible hypothesis, every revision of the frequency of planets around nearby stars seems to indicate that their frequency increases rapidly as their mass decreases and any estimates of their frequencies seem to be revised towards higher values as the amount of data accumulates \citep[e.g.][]{otoole2009b,howard2010,mayor2011,wittenmyer2011}.

\subsection{Radial velocity data}

\begin{figure}
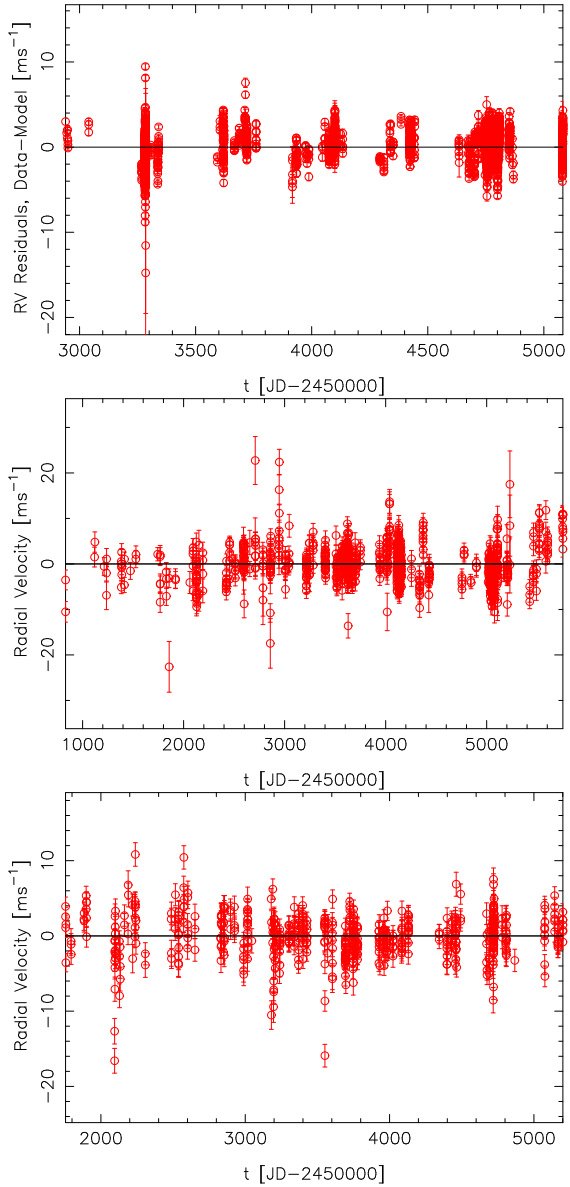

\center
\includegraphics[angle=-90,width=0.40\textwidth]{rvdist00_residc_rv_HD10700.ps}

\includegraphics[angle=-90,width=0.40\textwidth]{rvdist00_curvec_rv_HD10700e.ps}

\includegraphics[angle=-90,width=0.40\textwidth]{rvdist00_curvec_rv_HD10700f.ps}
\caption{HARPS (top), AAPS (middle), and HIRES (bottom) RVs with their respective mean estimates subtracted.}\label{velocity_fig}
\end{figure}

The HARPS spectra of HD 10700 were extracted from the ESO archive in a wavelength-calibrated form. This calibration was made using the HARPS Data Reduction Software (HARPS-DRS). After the spectral calibration, the RVs (Fig. \ref{velocity_fig}) were calculated using the cross-correlation function (CCF) technique presented in \citet{pepe2002}.

As the HARPS data ($N = 4864$, 205 epochs) contained some velocities that had significantly greater, i.e. more than 100 times greater, uncertainties than the HARPS velocities had in general, we simply neglected them and dropped them out of the data set prior to any further analyses. We also removed some spurious epochs from the set because they had velocities that differed by a few kms$^{-1}$ from the data mode. As a result, there were 4398 RV points (202 epochs) with a baseline of 2142 days. By visual inspection this data set indeed seemed ''flat``, as also described by \citet{pepe2011}, and is unlikely to contain periodic RV signals with amplitudes in excess of few ms$^{-1}$. The standard deviation of these data was found to be 1.7 ms$^{-1}$, which implies that there indeed could not be signals in the data in excess of roughly 2.0 ms$^{-1}$.

Calculating the Lomb-Scargle periodogram \citep{lomb1976,scargle1982} of the HARPS data revealed that this data set contains a ''jungle of peaks`` (Fig. \ref{HARPS_periodogram}, top panel) in excess of the analytical 0.1\% false alarm probabilities (FAPs). The reason is that the noise in this ''raw`` velocity data is probably not Gaussian nor white and thus violates the assumptions underlying the periodogram analyses. From this periodogram, it is thus difficult to interpret the significance of the corresponding peaks.

\begin{figure}
\center
\includegraphics[angle=-90,width=0.49\textwidth]{tauceti_raw.ps}

\includegraphics[angle=-90,width=0.49\textwidth]{tauceti_win.ps}
\caption{Lomb-Scargle periodogram (top) with 0.1\%, 1\%, and 10\% FAPs and the window function (bottom) of the HARPS data.}\label{HARPS_periodogram}
\end{figure}

The 978 AAPS RVs have a similar character to the HARPS data (Fig. \ref{velocity_fig}, middle panel). This data set, too, can be described as being flat and no periodic signals have been reported by the AAPS group despite an extensive baseline of the time-series of 4923 days \citep{wittenmyer2011b}. This data does not have such clear annual gaps as the HARPS data (Fig. \ref{velocity_fig}) and deviates from the mean by approximately 5.0 ms$^{-1}$ on average.

The smallest set of RVs was that measured using the HIRES (Fig. \ref{velocity_fig}, bottom panel). The 567 HIRES RV measurements have a baseline of 3446 days and nothing has been reported about this data set in the literature. The velocities deviate roughly 2.9 ms$^{-1}$ from the mean and do not have significant gaps apart from relatively narrow annual gaps corresponding to the visibility of HD 10700 from the Keck telescope's Northern location in Hawaii.

\section{Statistical analysis and modelling}\label{sec:modelling}

We modelled the RVs with a statistical model that contains three additive terms. These terms are (1) the superposition of $k$ Keplerian signals and a reference velocity, (2) Gaussian noise with two components, namely the instrument noise and all the excess noise in the measurements, and (3) an autoregressive (AR) and/or moving average (MA) term that accounts for the possible correlations between the subsequent velocities and/or their errors. The MA term accounts for the dependence of a measurement on the deviation of the last measurement from the mean -- this term corresponds effectively to binning measurements made within a certain timescale in order to remove variations within that timescale. While the first part of this model is described in detail in \citet{tuomi2011} and \citet{tuomi2011b}, we describe the rest of our statistical models and modelling approaches in the following subsections.

\subsection{Posterior samplings}

We analysed each model in our collection of candidate models using the adaptive Metropolis algorithm \citep{haario2001}. This algorithm works well for typical models of RVs and can be used to receive statistically representative samples from the posterior densities of the parameters of these models and with relatively little computational cost \citep{tuomi2011,tuomi2012,tuomi2011b}. The proposal density of the adaptive Metropolis algorithm is a multivariate Gaussian density, which poses problems and possibly slows down the convergence rate of the chains when the parameter posterior density is non-Gaussian with high non-linear correlations between the parameters. We overcome this problem by simply increasing the chain lengths sufficiently in each sampling to ensure that we obtain statistically representative samples. Typically, samples of roughly $10^{5}-10^{6}$, depending on the dimension of the parameter vector, are sufficient when the posterior density can be approximated as a multivariate Gaussian density. However, when there are several Keplerian signals in the model and the posterior density necessarily has non-linear correlations, we increase the chain lengths by factors of up to 100 to ensure that the samples we obtain are statistically representative.

The fact that the adaptive Metropolis algorithm is not exactly Markovian means that the sample drawn from the posterior density might not be close enough to the actual posterior density to draw conclusions on the parameters, and, on the model probabilities. We approximate these probabilities using the truncated posterior mixture (TPM) estimate of \citet{tuomi2012b} that requires a statistically representative sample from the posterior. For this reason, when the chains we calculate have converged close to the posterior density and are found to move randomly around in the vicinity of the maximum \emph{a posteriori} (MAP) estimate, we fixed the proposal density of the chain to its current value. This essentially makes this algorithm equivalent to the Metropolis-Hastings Markov chain Monte Carlo (MCMC) algorithm \citep{metropolis1953,hastings1970}. This is the same procedure that has been used in e.g. \citet{tuomi2012}.

We estimate all the parameters simply by using the MAP estimates and the 99\% Bayesian credibility sets \citep[as defined in e.g.][]{tuomi2009}, i.e. 99\% credibility intervals in a single dimension. For a definition of the MAP estimates, and the corresponding caveats of relying on point estimates in the first place, we refer the interested reader to any introductory text of Bayesian statistics \citep[e.g.][]{berger1980}.

\subsection{1st order AR model}

To be able to use autoregression in the statistical modelling and analysis, we arrange the RVs such that for epochs $t_{i}$ and $t_{j}$, $t_{i}<t_{j}$ if $i<j$, i.e. put the measurements in chronological order. Now, the measurement ($r_{i}$) at epoch $t_{i}$ depends on that at $t_{i-1}$ according to
\begin{eqnarray}\label{AR1_model}
  && r_{i} = f_{k}(t_{i}) + \epsilon_{i} + \epsilon_{J}, \textrm{ for } i=1 \nonumber\\
  && r_{i} = f_{k}(t_{i}) + \phi_{i,i-1} r_{i-1} + \epsilon_{i} + \epsilon_{J}, \textrm{ for } i>1 ,
\end{eqnarray}
where function $f_{k}$ is the superposition of $k$ Keplerian signals with an additive constant parameter $\gamma_{l}$, corresponding to the reference velocity of the $l$th telescope-instrument combination, $\epsilon_{i}$ is a Gaussian random variable with a zero mean and known variance corresponding to the estimated instrument noise $\sigma_{i}^{2}$ at $t_{i}$, the Gaussian random variable $\epsilon_{J}$ with zero mean and unknown variance ($\sigma_{J}^{2}$) represents all the excess (white) noise in the data.

The function $\phi_{i,j}$ describes the magnitude of the correlation between two subsequent measurements made at $t_{i}$ and $t_{j}$. We assume that the correlation described by this function depends on the time difference between the two consequent observations and write it as
\begin{equation}\label{correlation_function}
   \phi_{i,j} = \phi_{i} \exp \big[ \alpha (t_{j} - t_{i}) \big] ,
\end{equation}
where the positive numbers $\phi_{j}$, for all $j$, and $\alpha$ are free parameters of the model. It can be seen that the function $\phi_{i,j}$ decreases exponentially as the time difference between the two consequent epochs increases. Also, when $|\phi_{j}| < 1$ for all $j$ and $i > j$, its value is always less than unity, which makes the model stationary.

\subsection{1st order MA model}

To apply the MA model, we arrange the RVs again in chronological order. The RV measurement made at epoch $t_{i}$ is then modelled as
\begin{eqnarray}\label{MA1_model}
  && r_{i} = f_{k}(t_{i}) + \epsilon_{i} + \epsilon_{J}, \textrm{ for } i=1 \nonumber\\
  && r_{i} = f_{k}(t_{i}) + \omega_{i,i-1} (\epsilon_{i-1} + \epsilon_{J}) + \epsilon_{i} + \epsilon_{J}, \textrm{ for } i>1 ,
\end{eqnarray}
The MA part is defined using the function $\omega_{i,j}$ multiplied by the deviation of the previous observation from $f_{k}$. This function has the same form as $\phi_{i,j}$, with parameters $\omega_{j}$ and $\beta$ instead of $\phi_{j}$ and $\alpha$ in Eq. (\ref{correlation_function}), because we expect the dependence of the $i$th measurement on the deviation of the previous one to decrease as a function of the time difference of the corresponding measurements. 

\subsection{General ARMA model}

It may be the case in practice that taking into account the correlations between measurements at epochs $t_{i}$ and $t_{i-1}$ is not sufficient but a better model can be constructed by taking into account the correlations between the measurement at $t_{i}$ and those at $t_{i-j}, j = 1, ..., q$, where $q<i$. We write the corresponding $q$th order AR model as
\begin{eqnarray}\label{ARq_model}
  && r_{i} = f_{k}(t_{i}) + \epsilon_{i} + \epsilon_{J}, \textrm{ for } i=1 \nonumber\\
  && r_{i} = f_{k}(t_{i}) + \epsilon_{i} + \epsilon_{J} + \sum_{j=1}^{q} \phi_{j,i-j} r_{i-j}, \textrm{ for } i>1 .
\end{eqnarray}

Similarly, the $p$th order MA model is defined as
\begin{eqnarray}\label{MAp_model}
  && r_{i} = f_{k}(t_{i}) + \epsilon_{i} + \epsilon_{J}, \textrm{ for } i=1 \nonumber\\
  && r_{i} = f_{k}(t_{i}) + \epsilon_{i} + \epsilon_{J} + \sum_{j=1}^{p} \omega_{j,i-j} (\epsilon_{i-j} + \epsilon_{J}), \textrm{ for } i>1 .
\end{eqnarray}

The general ARMA model is then written by including both AR and MA terms in the statistical model together with the function $f$ and the additive random variables representing the white noise components in the data. We denote this general model as AR($q$)MA($p$).

\subsection{White noise models}

To determine the shape of the additive white noise in our statistical models, i.e. the distribution of the sum $\epsilon_{i} + \epsilon_{J}$, we compare some different distributions by relaxing the assumption that these two random variables (and consequently the sum) have Gaussian probability distributions. The Gaussian distribution is written simply as
\begin{equation}\label{gaussian}
  \mathcal{N}(\mu, \sigma^{2}) = \frac{1}{\sqrt{2\pi \sigma^{2}}} \exp \bigg\{ - \frac{(x - \mu)^{2}}{2 \sigma^{2}} \bigg\} ,
\end{equation}
and is well known for its property that independent random variables $X_{1} \sim \mathcal{N}(\mu_{1}, \sigma_{1}^{2})$ and $X_{2} \sim \mathcal{N}(\mu_{2}, \sigma_{2}^{2})$ satisfy $X_{1} + X_{2} \sim \mathcal{N}(\mu_{1} + \mu_{2}, \sigma_{1}^{2} + \sigma_{2}^{2})$.

The first alternative distribution we use is the Cauchy distribution $\mathcal{C}(x | \mu, \gamma) = \mathcal{C}(\mu, \gamma)$ defined as
\begin{equation}\label{cauchy}
  \mathcal{C}(\mu, \gamma) = \frac{1}{\pi} \Bigg[ \frac{\gamma}{(x - \mu)^{2} + \gamma^{2}} \Bigg] ,
\end{equation}
which has longer tails than the Gaussian distribution and satisfies the convenient property that for independent $X_{1} \sim \mathcal{C}(\mu_{1}, \gamma_{1})$ and $X_{2} \sim \mathcal{C}(\mu_{2}, \gamma_{2})$ it holds that $X_{1} + X_{2} \sim \mathcal{C}(\mu_{1} + \mu_{2}, \gamma_{1} + \gamma_{2})$. This distribution is selected to see if the noise is dominated by outliers that cannot be explained by the relatively short and ''light`` tails of the Gaussian distribution.

\begin{figure}
\center
\includegraphics[angle=270,width=0.40\textwidth]{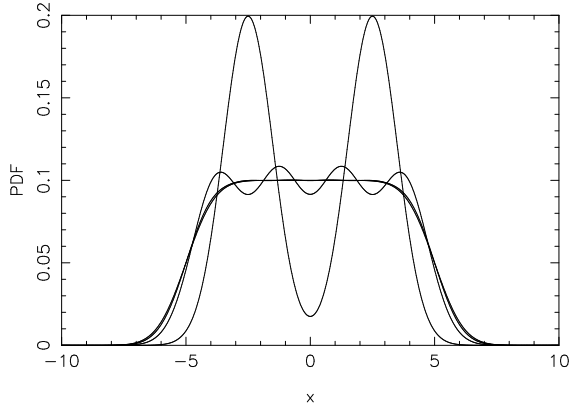}
\caption{Estimated probability distribution $\mathcal{S}(x | a, \sigma^{2})$ in Eq. (\ref{convolution}),  with $n = 1, ...,6$ (the corresponding functions have $2n$ maxima). The parameters are selected as $a = 5$ and $\sigma^{2}=1$.} \label{numerical_convolutions}
\end{figure}

Our second alternative model assumes that $\epsilon_{J} \sim \mathcal{U}(-a,a) \ast \mathcal{N}(0, \sigma_{J}^{2})$, where $\mathcal{U} (-a,a)$ is a uniform distribution of interval [$-a,a$], and $\epsilon_{i} \sim \mathcal{N}(0, \sigma_{i}^{2})$ are independent. In this case, their sum is distributed according to the convolution of the densities which we approximate as
\begin{equation}\label{convolution}
  \mathcal{S}(x | a, \sigma^{2}) = \lim_{n \rightarrow \infty} \frac{1}{2n} \sum_{k=0}^{2n-1} \mathcal{N} \bigg(x \bigg| \frac{a}{2n} [1-2n+2k], \sigma^{2} \bigg) ,
\end{equation}
where $\sigma^{2} = \sigma_{i}^{2} + \sigma_{J}^{2}$. In practice, when the values of $a$ and $\sigma$ are of the same magnitude, approximating the above infinite sum with a choice of $n=6$ provides an accurate estimate for the distribution. This distribution has a ``flat`` maximum but tails according to the Gaussian function, as seen in Fig. \ref{numerical_convolutions} for $a=5$ and $\sigma=1$. Even when parameter $a$ is five times greater than $\sigma$, the approximation converges very rapidly and is an accurate description of the convolution for $n = 3$ -- for obvious reasons decreasing $a$ or increasing $\sigma$ or $n$ improves this accuracy. As seen in Fig. \ref{numerical_convolutions}, the curves for $n = 3, ..., 6$ are practically indistinguishable from one another. We choose this model to investigate if the peak of the white noise component is not as sharp as in the Gaussian (or indeed in the Cauchy) model because a range of values at the vicinity of zero have roughly equal probabilities.

To summarise the above, our white noise models consist of (1) a Gaussian distribution because -- according to our knowledge --  it is the only distribution that has been used to model the measurement noise when analysing RV data, (2) a Cauchy distribution with longer tails, and (3) a flatter distribution that accounts for jitter with small deviations from the mean equally likely regardless of their exact magnitude. While these distributions are by no means a comprehensive collection of white noise models, we expect the comparisons of these models to provide information on the overall shape of the white noise component caused by the telescope-instrument combination and the stellar surface.

\subsection{Prior choice}

Because we take advantage of Bayesian inference, the choice of priors needs to be addressed briefly. Essentially, we use the same prior probability densities and prior probabilities that were used in \citet{tuomi2012}, there with the much more restricted data set of HD 10180. In particular, we chose $\pi(e) \propto \mathcal{N}(0, 0.3^{2})$, with the corresponding scaling, which penalises eccentricities more as they get closer to unity, yet, allows them if the higher eccentricities are needed to better describe the data. Because it is a scale invariant parameter, we adopt the logarithm of the period as a parameter and use a uniform prior with cutoffs $T_{min}$ and $T_{max}$ corresponding to one day and $10 T_{obs}$, respectively, where $T_{obs}$ is the baseline of the data analysed. We choose $T_{max} > T_{obs}$ because signals with periods in excess of the data baseline can be detected in RV data sets \citep{tuomi2009b}.

The noise models have additional parameters as well, and their priors were chosen as follows. The parameters $\phi_{j}$ (and $\omega_{j}$) were set to have uniform priors in the interval [-1, 1], for all $j$, to allow both positive and negative correlations to occur. Although, we expected these parameters to receive mostly positive values that were at least consistent with zero given their uncertainties. The parameter $\alpha$ ($\beta$), which determines the time-scale of the AR (MA) effect in the noise, was chosen to have a uniform prior in the logarithmic scale, i.e. the Jeffreys prior, with cutoffs at $\alpha_{min} = 1$ min$^{-1}$ and $\alpha_{max} = 1$ year$^{-1}$. Though we expected this effect to take place on the time-scale of hours, we chose a wider interval limited by the minimum time-difference between two subsequent measurements ($\sim$ 1 min) and a maximum of one year to account for possible noise correlations on longer time-scales as well.

We choose the prior probabilities, $P(\mathcal{M}_{i})$, of our noise models ($\mathcal{M}_{i}$) to be equal. However, when comparing models with different numbers of Keplerian signals we do not assume equal prior probabilities but set them such that they favour the model with one less signal by a factor of two. These priors have been applied in e.g. \citet{tuomi2012}.

Given the unprecedented amount of data, we expect the priors to be overwhelmed in the case of HD 10700 velocities and that they do not have a detectable effect on the results. For this reason, we do not test the dependence of our results on the choice of prior densities.

\section{Model selection}\label{sec:noise_selection}

We take advantage of the Bayesian model selection framework, in which each model is equipped with a number that describes its relative posterior probability given the measurements. The fact that this probability is relative means that it only tells how probable it is that the measurements, as random variables, have been drawn from the random process described by the model instead of being drawn from those described by the other models in the model set. Bayesian model selection has been used in determining the most probable number of planetary signals in the RV data \citep[e.g.][]{gregory2005,gregory2007a,gregory2007b,gregory2011,ford2007,tuomi2009,tuomi2011,tuomi2012,tuomi2011b} but it can be readily applied to any other model comparison problem because of its generality.

We compare the models by calculating the posterior probabilities as
\begin{equation}\label{model_probability}
  P(\mathcal{M}_{i} | m) = \frac{P(m | \mathcal{M}_{i})P(\mathcal{M}_{i})}{\sum_{j=1}^{k} P(m | \mathcal{M}_{j})P(\mathcal{M}_{j})} ,
\end{equation}
where $P(m | \mathcal{M}_{i})$ are the marginal integrals whose values need to be calculated for each model given the measurements $m$. We estimate these integrals using the TPM estimate but verify the results using the simple Akaike information criterion \citep{akaike1973} for small sample size \citep[AICc, see e.g.][]{burnham2002} because the data sets we analyse are large enough so that the number of data points well exceeds the number of model parameters.

To ensure that the TPM estimates we obtain are trustworthy, we perform several samplings for each data set and statistical model with different initial states. We perform three such samplings to check that the obtained TPM estimates are consistent. If we find that the sample sizes are insufficient, we typically increase them by a factor of 10 and again obtain three independent samples. We found that even when there were several Keplerian signals in the model, sample sizes of the order of $10^{7}$ were sufficient and independent samplings yielded TPM estimates within roughly 0.05 from one another in the log-scale. We note that the parameters $\lambda$ and $h$ in the algorithm of the TPM estimate \citep{tuomi2012b} were chosen to be 10$^{-5}$ and 10$^{5}$, respectively, because the chain members $\theta_{n}$ and $\theta_{n+h}$ became independent for roughly $h \approx 10^{3} - 10^{4}$ and parameter $\lambda$ of $10^{-5}$ was found to converge rather rapidly but still provide TPM estimates within roughly 0.05 from one another in the log-scale.

In practice, we first compare different AR and MA models to find out the most reliable description of the nature and amount of autoregression and noise correlation in the data. After that, we compare the different models for the remaining white noise in the data. We perform the comparisons by using the HARPS RVs of HD 10700. We generate a total of fourteen data sets from the 4398 HARPS RVs by adding Keplerian signals to the time-series to see which ones of the AR and MA models yield the correct RV amplitudes for these artificial signals. Out of the models that yield results consistent with the parameters of the artificially added signals, we select a model that has the greatest posterior probability according to Eq. (\ref{model_probability}). We do not simply choose blindly the best model according to this Eq. but require primarily that the model yields results consistent with the Keplerian signals introduced into the data. The reason for this choice is that it is possible that a signal, or at least part of it, gets interpreted as being a consequence of significant autoregression in the data, or is generated by pure noise by some other means. Therefore, especially with respect to the AR models, we analyse the reliability of the different noise models carefully.

After having found the best descriptions out of the AR and MA models, especially the ones that do not result in biases in the artificial signals, we perform a comparison of our white noise models.

\section{Signal detection criteria}\label{sec:signal_detection}

Before analysing the RV data with and without artificially added signals, we discuss briefly the requirements for a positive detection of a periodic signal in the data. Because these criteria are essential in the detection of weak signals in, not only RVs, but in any measurements aiming at detecting periodic signals, we describe our criteria in this section.

The regular approach to detections of Keplerian signals in RVs is based on the Lomb-Scargle periodogram of residuals that are assumed to have a Gaussian distribution \citep{lomb1976,scargle1982}. While we studied the periodogram powers in our analyses, we do not use them as an indication of whether there is a periodic signal in the data or not. The reason for this choice is that calculating the periodogram of model residuals assumes the remaining noise is Gaussian and that there are no additional signals -- if there are, the assumptions are violated and the test cannot be considered trustworthy \citep[see e.g.][]{tuomi2012}.

The analytical detection threshold of \citet{tuomi2009b} can be used to receive a rough estimate for the detectability of various signals in a given data set. According to this criterion, a periodic signal with period $P$ can be detected if the square of its amplitude exceeds the threshold given by
\begin{eqnarray}\label{threshold}
  && K_{T}^{2} = \frac{9.22 \sigma^{2}}{N} \bigg[ f_{c}^{2}(\psi) + f_{s}^{2}(\psi) \bigg] \textrm{, where} \\
  && f_{c} = 2\big[ 1 - \cos \pi \psi \big]^{-1} \textrm{ if } \psi < 1 \textrm{ and unity otherwise} ,\nonumber\\
  && f_{s} = \big[ \sin \pi \psi \big]^{-1} \textrm{ if } \psi < 0.5 \textrm{ and unity otherwise} , \nonumber
\end{eqnarray}
where $N$ is the number of measurements, $\sigma$ is the average noise level of the data, and $\psi = T P^{-1}$, where $T$ is the baseline of the data. This criterion is only a very rough estimate because it does not take into account the various effects that data sampling might have on the detectability of signals. However, it does take into account the ratio of the data baseline and the period of the signal, which means that the criterion is applicable even in cases where the period of the signal exceeds the data baseline.

Using the criterion in Eq. (\ref{threshold}), we find that signals with periods less than roughly 1000 days can be detected in the HARPS RVs of HD 10700 if their amplitudes exceed 0.1 ms$^{-1}$. In practice, the amplitude of a signal has to be significantly above this threshold, i.e. in excess of the 99\% Bayesian credibility interval. While this threshold appears to be very low, it results from the fact that the data set contains a large number of high-precision RVs.

Because the threshold in Eq. (\ref{threshold}) is only a rough approximation, we use more robust criteria to determine whether signals are present in a data set. We  say that $k+1$ signals are detected if (1) the posterior probability of a model with $k+1$ signals is at least 150 times greater than that of a model with $k$ signals \citep{kass1995,feroz2011,tuomi2012}, if (2) the RV amplitudes of all signals are statistically significantly greater than zero, and if (3) the periods of all signals are well constrained from above and below \citep{tuomi2012}. We adopt these criteria but require also that the signal amplitude exceeds the threshold presented in Eq. (\ref{threshold}).

\section{Artificial signals}\label{sec:artificial}

We added twelve different Keplerian signals to the HARPS RV data of HD 10700 to see which models could extract their parameters from these artificial data sets the most accurately. These signals were set to have periods of 20, 50, 100, and 200 days and amplitudes of 1, 2, and 5 ms$^{-1}$, which resulted in a collection of twelve data sets. We also generated two additional sets by adding a 200 days periodicity with amplitudes of 0.5 and 0.3 ms$^{-1}$ to see whether such extremely low-amplitude signals could be retrieved from the data. We chose the 200 days period because the power spectrum of the HARPS velocities had the fewest peaks between roughly 150 and 300 days (Fig. \ref{HARPS_periodogram}). We state that a signal is detected reliably if (1) the detection criteria in the previous section are satisfied and if (2) the 99\% Bayesian credibility sets \citep[BCSs; as defined in e.g.][]{tuomi2009}, i.e. intervals in one dimension, of the period and amplitude parameters contain the correct values of the added artificial signals.

\begin{table*}
\center
\caption{Artificial (first column) and retrieved (subsequent columns) signals (their MAP parameter estimates) in terms of parameters $K$, $P$, and $e$ using a model without any of the AR or MA components (0) and with selected AR or MA models for each data set. Signals detected reliably, i.e. whose 99\% Bayesian credibility intervals contain the parameter values of the artificially added signals, are emphasised using boldface.}\label{model_performance}
\begin{tabular}{lcccccccc}
\hline \hline
Signal & 0 & AR(1) & MA(1) & AR(3) & MA(3) & MA(5) & MA(7) & MA(10) \\
\hline
$K = 5$ [ms$^{-1}$] & 5.50 & 2.87 & 5.49 & 2.03 & 5.32 & 5.36 & 5.39 & 5.38 \\
$P = 20$ [days] & 19.99 & 19.99 & 19.99 & 19.99 & \textbf{20.00} & \textbf{20.00} & \textbf{20.00} & \textbf{20.00} \\
$e = 0$ & 0.05 & 0.06 & 0.05 & \textbf{0.00} & \textbf{0.01} & \textbf{0.03} & \textbf{0.00} & \textbf{0.02} \\
\hline
$K = 2$ [ms$^{-1}$] & 2.60 & 1.25 & 2.57 & 0.75 & 2.39 & 2.41 & 2.37 & 2.40 \\
$P = 20$ [days] & 19.98 & 19.98 & 19.98 & 19.97 & \textbf{19.99} & \textbf{20.00} & \textbf{20.00} & \textbf{20.00} \\
$e = 0$ & 0.08 & 0.08 & \textbf{0.07} & 0.10 & \textbf{0.08} & \textbf{0.02} & \textbf{0.05} & \textbf{0.03} \\
\hline
$K = 1$ [ms$^{-1}$] & 1.68 & 0.80 & 1.65 & 0.44 & 1.41 & 1.39 & 1.39 & 1.40 \\
$P = 20$ [days] & 19.96 & 19.97 & 19.97 & 19.96 & \textbf{19.98} & \textbf{20.00} & \textbf{20.00} & \textbf{20.01} \\
$e = 0$ & 0.11 & \textbf{0.06} & \textbf{0.10} & \textbf{0.01} & \textbf{0.07} & \textbf{0.07} & \textbf{0.03} & \textbf{0.03} \\
\hline
$K = 5$ [ms$^{-1}$] & 5.32 & 2.63 & 5.33 & 1.78 & 5.30 & 5.31 & 5.27 & 5.31 \\
$P = 50$ [days] & 50.08 & 50.08 & 50.07 & 50.06 & 50.05 & 50.05 & 50.04 & 50.05 \\
$e = 0$ & 0.07 & 0.07 & 0.07 & \textbf{0.00} & 0.05 & \textbf{0.04} & \textbf{0.00} & \textbf{0.00} \\
\hline
$K = 2$ [ms$^{-1}$] & 2.42 & 1.13 & 2.45 & 0.65 & 2.40 & 2.37 & 2.33 & 2.33 \\
$P = 50$ [days] & 50.12 & 50.13 & 50.14 & 50.17 & 50.11 & 50.09 & 50.09 & 50.11 \\
$e = 0$ & 0.14 & 0.13 & 0.13 & \textbf{0.03} & 0.11 & \textbf{0.08} & \textbf{0.07} & \textbf{0.07} \\
\hline
$K = 1$ [ms$^{-1}$] & 1.58 & 0.72 & 1.56 & 0.39 & 1.41 & 1.37 & 1.36 & 1.36 \\
$P = 50$ [days] & 50.21 & 50.21 & 50.20 & 50.21 & 50.16 & 50.16 & 50.17 & 50.19 \\
$e = 0$ & 0.25 & 0.22 & 0.20 & \textbf{0.21} & \textbf{0.15} & \textbf{0.11} & \textbf{0.10} & \textbf{0.05} \\
\hline
$K = 5$ [ms$^{-1}$] & 5.46 & 2.64 & 5.45 & 1.76 & 5.39 & 5.40 & 5.36 & 5.35 \\
$P = 100$ [days] & \textbf{99.98} & 99.85 & \textbf{99.92} & 100.07 & \textbf{99.96} & \textbf{99.99} & \textbf{99.98} & \textbf{99.99} \\
$e = 0$ & 0.09 & \textbf{0.00} & 0.07 & 0.04 & 0.05 & \textbf{0.04} & \textbf{0.05} & \textbf{0.03} \\
\hline
$K = 2$ [ms$^{-1}$] & 2.67 & 1.30 & 2.58 & 0.68 & 2.43 & 2.36 & 2.39 & 2.36 \\
$P = 100$ [days] & 100.47 & 100.43 & 100.38 & \textbf{100.02} & \textbf{100.00} & \textbf{99.99} & \textbf{100.02} & \textbf{100.00} \\
$e = 0$ & 0.30 & 0.30 & 0.30 & \textbf{0.05} & 0.16 & \textbf{0.11} & \textbf{0.10} & \textbf{0.10} \\
\hline
$K = 1$ [ms$^{-1}$] & 1.72 & 0.80 & 1.63 & 0.44 & 1.48 & 1.41 & 1.42 & 1.40 \\
$P = 100$ [days] & 100.48 & 100.45 & 100.41 & \textbf{100.44} & \textbf{100.06} & \textbf{100.11} & \textbf{100.07} & \textbf{100.13} \\
$e = 0$ & 0.30 & 0.29 & 0.29 & \textbf{0.27} & \textbf{0.27} & \textbf{0.24} & \textbf{0.17} & \textbf{0.19} \\
\hline
$K = 5$ [ms$^{-1}$] & \textbf{4.84} & 2.34 & \textbf{4.86} & 1.46 & \textbf{4.91} & \textbf{5.00} & \textbf{5.03} & \textbf{5.03} \\
$P = 200$ [days] & 200.74 & 200.69 & 200.62 & 200.66 & \textbf{200.34} & \textbf{200.15} & \textbf{200.14} & \textbf{200.09} \\
$e = 0$ & 0.08 & 0.06 & 0.07 & \textbf{0.02} & 0.07 & 0.07 & 0.07 & 0.07 \\
\hline
$K = 2$ [ms$^{-1}$] & \textbf{1.95} & 0.94 & \textbf{1.96} & 0.59 & \textbf{2.03} & \textbf{2.00} & \textbf{2.05} & \textbf{2.09} \\
$P = 200$ [days] & 201.97 & 201.73 & 201.71 & \textbf{201.33} & \textbf{199.92} & \textbf{201.04} & \textbf{200.13} & \textbf{199.84} \\
$e = 0$ & 0.14 & \textbf{0.12} & 0.14 & \textbf{0.09} & \textbf{0.16} & 0.13 & \textbf{0.12} & 0.12 \\
\hline
$K = 1$ [ms$^{-1}$] & \textbf{1.13} & 0.54 & \textbf{1.06} & 0.33 & \textbf{1.05} & \textbf{1.10} & \textbf{1.11} & \textbf{1.12} \\
$P = 200$ [days] & 203.55 & 203.18 & 203.55 & \textbf{202.14} & \textbf{201.92} & \textbf{201.68} & \textbf{200.57} & 198.06 \\
$e = 0$ & 0.20 & \textbf{0.14} & 0.15 & \textbf{0.12} & \textbf{0.14} & \textbf{0.15} & \textbf{0.15} & \textbf{0.14} \\
\hline
$K = 0.5$ [ms$^{-1}$] & - & - & - & - & - & \textbf{0.66} & \textbf{0.64} & \textbf{0.65} \\
$P = 200$ [days] & - & - & - & - & - & \textbf{202.08} & \textbf{201.97} & \textbf{201.36} \\
$e = 0$ & - & - & - & - & - & \textbf{0.09} & \textbf{0.13} & \textbf{0.11} \\
\hline
$K = 0.3$ [ms$^{-1}$] & - & - & - & - & - & \textbf{0.50} & \textbf{0.48} & \textbf{0.48} \\
$P = 200$ [days] & - & - & - & - & - & \textbf{201.84} & \textbf{202.62} & \textbf{203.16} \\
$e = 0$ & - & - & - & - & - & \textbf{0.17} & \textbf{0.11} & \textbf{0.09} \\
\hline \hline
\end{tabular}
\end{table*}

According to the results presented in Table \ref{model_performance}, the pure white noise model and both first order AR and MA models could not be used to determine the parameter values of the artificial signals reliably. This was found to be the case even with the signals with amplitudes as high as 5 ms$^{-1}$ that should be detectable from high-precision data easily, especially, given the large number of data points. We also found this to be the case for higher order AR models that yielded severe biases in the signals because the signals were interpreted, in part, as noise-related correlations in the data. Therefore, despite the addition of AR components to the noise model improving the goodness of the model, we do not consider the AR models trustworthy for our purposes. According to the results in Table \ref{model_performance}, the AR models underestimate the signal amplitudes significantly.

The MA models were found to be reasonably reliable in quantifying the properties of the signals in the data. While they overestimated slightly the amplitudes of the 20, 50, and 100 day signals, they were accurate for the 200 day signal though again somewhat overestimated the signal amplitudes when recovering the 0.3 or 0.5 ms$^{-1}$ injected signals. Also, apart from the 50 day signal, the MA models of order seven and ten yielded the best results for periods and eccentricities of the injected signals.

The reason the 50 day signal could not be extracted correctly and the fact that the MA models, even the most accurate (i.e. that have the greatest posterior probabilities) seventh and tenth order ones, yielded biased estimates for the amplitudes of the 20 and 100 day day signals warrants an explanation. We are essentially studying the properties of the noise in the HARPS data of HD 10700. However, there is a possibility that the noise models we use lack some important features that impinges on the ability to relialbly recover injected signals. Another possibility is that there are already signals present in the data and we actually detect the superpositions of these real signals and the artificial ones. If this is the case, the above considerations depend on how much the existing signals affect the artificial ones. We expect that there are no significant signals at or around 200 days in the data because the 200 day signals were extracted the most accurately with the best MA models. However, in Section \ref{sec:harps_analysis} we show that there are genuine low-amplitude signals in the data near the periods that provided relatively poor recovery of signals and so the lack of complete recovery of injected signals is not surprising.

The artificial signals at 200 days with the lowest amplitudes received slightly biased amplitudes. The amplitudes of the recovered artificial signals were systematically around 0.15-0.20 ms$^{-1}$ greater than their real values. While these values were within the 99\% BCSs of the obtained estimates, this over-estimation is a rather awkward feature and implies that the models are not as good descriptions of the data as they should be. Yet, this might again be a caused by the fact that there are signals -- or their aliases -- at or around 200 days in the HARPS data that cause biases to our estimates.

In Table \ref{model_performance}, the estimates are only shown for models MA(5), MA(7), and MA(10), because the other models did not identify any significant periodicities at or around 200 days.

\begin{figure*}
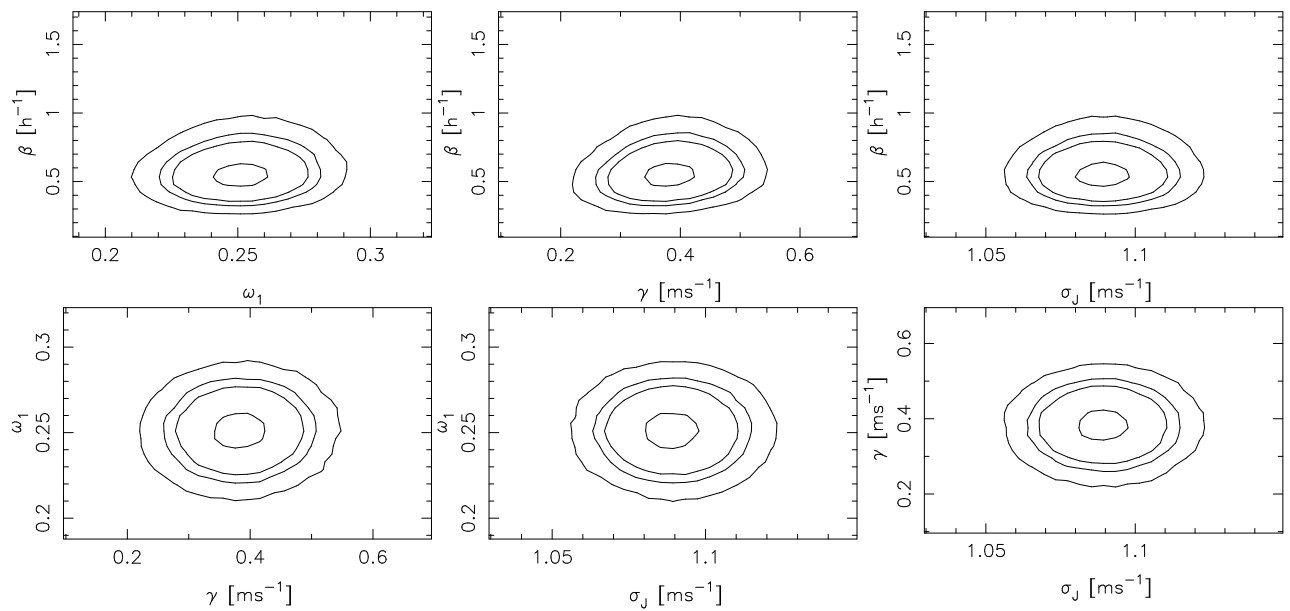

\center
\includegraphics[angle=270,width=0.30\textwidth]{rvdist00_rv_HD10700_co_alf0.ps}
\includegraphics[angle=270,width=0.30\textwidth]{rvdist00_rv_HD10700_co_alga.ps}
\includegraphics[angle=270,width=0.30\textwidth]{rvdist00_rv_HD10700_co_alsj.ps}

\includegraphics[angle=270,width=0.30\textwidth]{rvdist00_rv_HD10700_co_f0ga.ps}
\includegraphics[angle=270,width=0.30\textwidth]{rvdist00_rv_HD10700_co_f0sj.ps}
\includegraphics[angle=270,width=0.30\textwidth]{rvdist00_rv_HD10700_co_gasj.ps}
\caption{Equiprobability contours containing 50\%, 10\%, 5\%, and 1\% of the probability density of all the combinations of parameters $\beta$, $\omega_{1}$, $\gamma$, and $\sigma_{J}$ from a single Markov chain using a model without Keplerian signals and a MA(10) noise description.} \label{fig:contours}
\end{figure*}

When sampling the posterior density, we observed that the joint posterior density of the noise parameters and reference velocity was close-Gaussian in all the samplings we performed. Therefore, we are confident that the adaptive Metropolis algorithm enables fast convergence and enables us to draw statistically representative samples. We illustrate this by plotting the equiprobability contours of parameters $\beta$, $\omega_{1}$, $\gamma$, and $\sigma_{J}$ in Fig. \ref{fig:contours} as obtained using the HARPS data and a model without Keplerian signals.

\section{HARPS radial velocities of HD 10700}\label{sec:harps_analysis}

\subsection{Noise model}

We analysed the HARPS RVs using several different noise models. We chose the same AR or MA models as in Table \ref{model_performance} and calculated their posterior probabilities to compare their performances in explaining the data. We started with pure noise models without any Keplerian signals. As was the case with the datasets with injected artificial signals, the AR models (AR(1) and AR(3)), not to mention the pure white noise model, did not perform as well as the MA models (Table \ref{HARPS_noise}). According to our probabilities in Table \ref{HARPS_noise}, the MA(10) model was the best description of the data because it had greater posterior probability than the MA(7) model. It can be seen that adding components to the MA model improves its performance until there are roughly 5-10 components. Therefore, we did not try more components and continued analysing the data with the MA(10) noise model that had the greatest posterior probability.

We also tested the relative performance of the two different white-noise models, namely the Cauchy and the convolution of Gaussian and uniform density (C and G+U in Table \ref{HARPS_noise}, respectively), in explaining the HARPS velocities. According to our results, the Cauchy noise model does not perform well with respect to the Gaussian white noise model (Table \ref{HARPS_noise}). The uniform component, containing one extra parameter that penalises the probability of this model in accordance with the principle of parsimony, does not increase the performance of the noise model enough to receive a greater posterior probability than the Gaussian model. This indicates that the white noise component of the HARPS velocities has a shape that resembles the Gaussian density and can be modelled well using the density in Eq. (\ref{gaussian}) if the variance is written as the sum of the variances of the instrument noise ($\sigma_{i}^{2}$) and the excess noise ($\sigma_{J}^{2}$) for every measurement $r_{i}$. In practice this also means that the additional parameter $a$ in Eq. (\ref{convolution}) is statistically indistinguishable from zero in this case.

\begin{table}
\center
\caption{Relative posterior probabilities and log-Bayesian evidences for different noise models using the HARPS RVs. No Keplerian signals were included in the model. The noise models are Gaussian (G), Cauchy (C), Uniform (U), AR($q$), and MA($p$).}\label{HARPS_noise}
\begin{tabular}{lcc}
\hline \hline
Model & $P(\mathcal{M} | d)$ & $\ln P(d | \mathcal{M})$ \\
\hline
G & $\sim 10^{-673}$ & -8431.0 \\
G+AR(1) & $\sim 10^{-245}$ & -7446.0 \\
G+MA(1) & $\sim 10^{-245}$ & -7447.3 \\
G+AR(3) & 3.2$\times 10^{-53}$ & -7004.5 \\
G+MA(3) & 1.1$\times 10^{-48}$ & -6994.0 \\
G+MA(5) & 2.8$\times 10^{-14}$ & -6914.8 \\
G+MA(7) & 1.9$\times 10^{-9}$ & -6903.7 \\
G+MA(10) & 0.61 & -6884.1 \\
C+MA(10) & $\sim 10^{-167}$ & -7267.1 \\
G+U+MA(10) & 0.39 & -6884.5 \\
\hline \hline
\end{tabular}
\end{table}

Thus we proceed to model the noise in the HARPS velocities using the tenth order MA model and Gaussian white noise.

\subsection{Signals in the HARPS data}

After removing the MAP estimated MA(10) components of the noise from the data and calculating the Lomb-Scargle periodogram of the residuals, most of the peaks appearing in the ''raw`` velocity data (Fig. \ref{HARPS_periodogram}) seem to disappear from the power spectrum (Fig. \ref{periodograms}, top panel). However, there is one strong peak that exceeds the 1\% FAP at 35.3 days and four others that exceed the 10\% FAPs at 13.9, 20.1, 363, and 595 days. Since we did not perform binning of the data, the measurements likely contain more information than the binned data from which significant periodicities have not been found \citep{pepe2011}. Therefore, we added one Keplerian signal to our model and calculated its posterior probability to see if the strongest peaks in the periodogram were significant signals according to our criteria.

\begin{figure}
\center
\includegraphics[angle=-90,width=0.49\textwidth]{tauceti_0.ps}

\includegraphics[angle=-90,width=0.49\textwidth]{tauceti_1.ps}

\includegraphics[angle=-90,width=0.49\textwidth]{tauceti_2.ps}

\includegraphics[angle=-90,width=0.49\textwidth]{tauceti_3.ps}
\caption{Lomb-Scargle periodograms of the HARPS data residuals after removing moving average components from the noise (top panel) and removing the 35, 14, and 95 day signals (subsequent panels). The dotted, dashed, and dot-dashed lines indicate the analytical 0.1\%, 1\%, and 10\% FAPs, respectively.}\label{periodograms}
\end{figure}

The one-Keplerian model increased the model probability by a factor of 1.8$\times 10^{16}$ (Table \ref{HARPS_probabilities}), which makes the corresponding periodicity of 35 days significantly present in the data. However, the residuals of the one-Keplerian model showed an additional peak exceeding the 1\% FAP at 14 days (Fig. \ref{periodograms}, second panel), and we continued analysing the data with a two-Keplerian model. This model further increased the model probability by a factor of 9.6$\times 10^{10}$. Finally, the residuals of this two-Keplerian model contained one more signal at roughly 94 days that exceeded the 10\% FAP level (Fig. \ref{periodograms}, third panel). Again, the corresponding periodicity in a three-Keplerian model increased the model probability significantly by a factor of 1.2$\times 10^{17}$. After removing this signal there were no strong powers in the power spectrum (Fig. \ref{periodograms}, bottom panel) and the samplings of the parameter space of a four-Keplerian model failed to identify any additional significant periodicities. Therefore, there appear to be three Keplerian signals in the HARPS RVs of $\tau$ Ceti. The posterior probabilities in Table \ref{HARPS_probabilities} indicate that taking these signals into account improves the model very significantly, which implies that there is strong evidence for three periodic signatures in the $\tau$ Ceti data. The probabilities based on the simple AICc imply the same qualitative results (Table \ref{HARPS_probabilities}).

\begin{table}
\center
\caption{Relative posterior probabilities and log-Bayesian evidences of models $\mathcal{M}_{k}$ with $k = 0, ..., 3$ Keplerian signals, given the HARPS RVs.}\label{HARPS_probabilities}
\begin{tabular}{lcccc}
\hline \hline
$k$ & $P(\mathcal{M}_{k} | d)$ & $\ln P(d | \mathcal{M}_{k})$ & $P(\mathcal{M}_{k} | d)$ & $\ln P(d | \mathcal{M}_{k})$ \\
 & TPM & TPM & AICc & AICc \\
\hline
0 & 4.7$\times10^{-45}$ & -6884.1 & 1.0$\times10^{-39}$ & -6886.7 \\
1 & 8.4$\times10^{-29}$ & -6846.0 & 1.6$\times10^{-24}$ & -6851.1 \\
2 & 8.1$\times10^{-18}$ & -6820.0 & 7.4$\times10^{-14}$ & -6825.8 \\
3 & $\sim$ 1 & -6780.0 & $\sim$ 1 & -6794.9 \\
\hline \hline
\end{tabular}
\end{table}

\begin{figure}
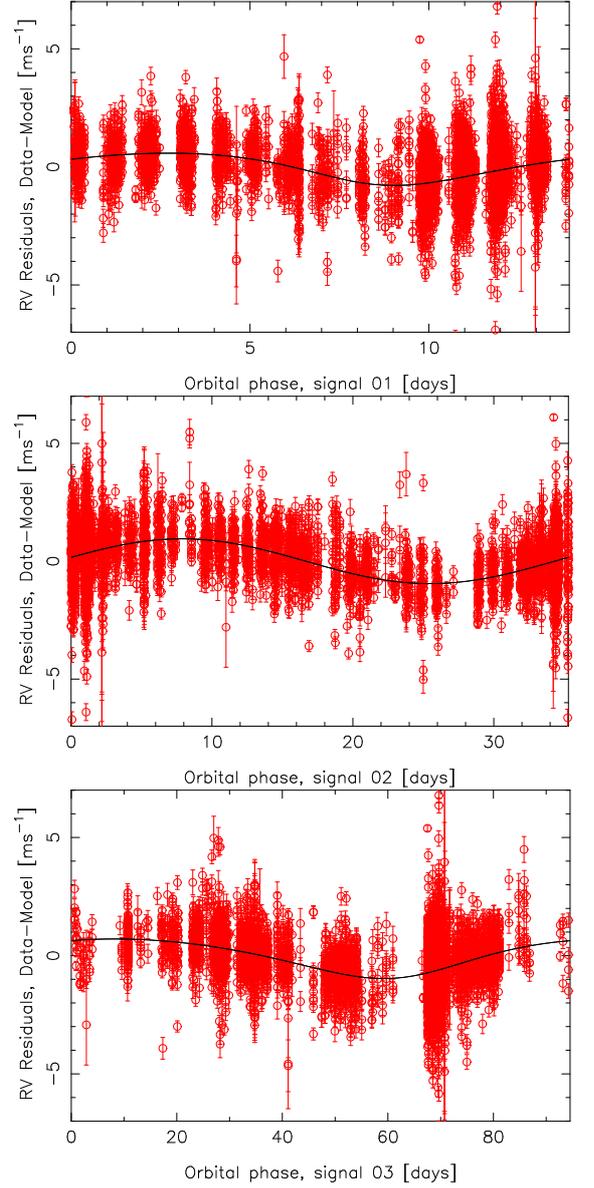

\center
\includegraphics[angle=-90,width=0.40\textwidth]{rvdist03_scresidc_rv_HD10700_1.ps}

\includegraphics[angle=-90,width=0.40\textwidth]{rvdist03_scresidc_rv_HD10700_2.ps}

\includegraphics[angle=-90,width=0.40\textwidth]{rvdist03_scresidc_rv_HD10700_3.ps}
\caption{Phase-folded signals of the three-Keplerian solution with the other two signals and the moving average component (MA(10)) of the noise removed.}\label{orbits}
\end{figure}

We show the MAP phase-folded orbits of our three-Keplerian solution in Fig. \ref{orbits}. While these plots are not visually very impressive, they do indicate that the large amount of data together with the improved modelling of the noise enables the detection of these signals. The RV amplitudes of all three signals were found to have MAP estimates below 1.0 ms$^{-1}$ level but they were still well constrained and differ from zero very significantly (Table \ref{3Kep_HARPS}).

\begin{table*}
\center
\caption{MAP estimates and the corresponding 99\% BCSs of the parameters of the three-Keplerian model for the HARPS RVs.}\label{3Kep_HARPS}
\begin{tabular}{lcccc}
\hline \hline
Parameter & HD 10700 b & HD 10700 c & HD 10700 d \\
\hline
$P$ [d] & 13.927 [13.901, 13.957] & 35.314 [35.142, 35.434] & 94.32 [93.62, 95.12] \\
$e$ & 0.17 [0, 0.41] & 0.11 [0, 0.28] & 0.21 [0, 0.42] \\
$K$ [ms$^{-1}$] & 0.60 [0.30, 1.02] & 0.89 [0.52, 1.32] & 0.89 [0.42, 1.22] \\
$\omega$ [rad] & 2.2 [0, 2$\pi$] & 2.8 [0, 2$\pi$] & 3.9 [0, 2$\pi$] \\
$M_{0}$ [rad] & 3.4 [0, 2$\pi$] & 3.9 [0, 2$\pi$] & 5.1 [0, 2$\pi$] \\
$m_{p} \sin i$ [M$_{\oplus}$] & 2.0 [1.0, 3.3] & 4.0 [2.2, 5.7] & 5.5 [2.6, 7.5] \\
$a$ [AU] & 0.106 [0.099, 0.110] & 0.197 [0.184, 0.204] & 0.379 [0.356, 0.393] \\
\hline
$\sigma_{J}$ [ms$^{-1}$] & 1.05 [1.01, 1.12] \\
\hline \hline
\end{tabular}
\end{table*}

With respect to the artificial data sets we analysed in the previous section, we suspect that the artificial signals at 50 days did not get detected reliably because there already were periodicities at 35 and 94 days in the data. The strongest periodicity in the data, namely at 35 days with an amplitude of roughly 1 ms$^{-1}$, is likely preventing the reliable detection of the artificial 50 days signals. While we still could extract them significantly out of the data, their amplitudes were biased, which is likely caused by the fact that the superposition of the artificial and real signals (and/or their aliases) was actually the one that we detected. To test this hypothesis, we analysed one of the data sets with injected signal ($K=1.0$ ms$^{-1}$, $P=50$ days) while simultaneously modelling the existing signals in the data. While we still obtained biased estimates for the parameters $P$ and $K$, taking into account all three periodicities at periods of approximately 14, 35, and 94 days (Table \ref{3Kep_HARPS}) enabled us to retrieve the injected 50 day signal correctly given the uncertainties of the parameters. This indicates that the existing periodicities might indeed cause biases to the process of recovering the artificial signals. We expect this is the reason the 20 and 100 days signals were poorly recovered from our artificial test data. However, there do not appear to be very strong periodicities around 200 days (Fig. \ref{periodograms}), which made it possible to extract the corresponding artificial signals from the data reasonably reliably given a sufficiently sophisticated noise model. Yet, even the artificial signals at 200 days received amplitudes that were systematically greater than the values used to generate them. This suggests that despite the fact that we could not detect additional signals in the HARPS data, there might still be some low-amplitude signals hidden in the measurement noise at longer periods.

We note that, with the samplings of the parameter space of models with more than one Keplerian signal, the non-linear correlations slow the convergence rate of the Markov chains. This effect arises because the multivariate Gaussian proposal density does not resemble the posterior very well. For this reason, we adopted a \emph{brute force} approach and simply increased the chain lengths suficiently to obtain samples that are statistically representative. In practice, we increased the chain lengths to up to few 10$^{7}$ to ensure that they had indeed converged to the posterior density.

The signals in the HARPS velocities are of low amplitude and it is therefore relevant to ask whether they can be retrieved from the HARPS data with different noise models. We tested the dependence of the extracted signals on the noise models by seeing whether the same probability maxima existed in the period space. The exact amount of MA components was found to impact the resuts little and we could retrieve the three signals using models with less (5) or more (12) MA components. With the MA(3) model, however, the period space was found to contain much more maxima likely arising from noise correlations that were not accounted for in the model. We could also obtain two of the signals (at periods of 14 and 35 days) by using an AR(5)MA(5) model and observed a maximum periodogram power in the residual periodogram at 95 days but could not constrain the corresponding signal by samplings. We could also detect the three signals by using the ''flat Gaussian`` likelihood model in Eq. (\ref{convolution}). These results indicate that the signals in the HARPS data are rather independent of the exact noise model.

\subsection{Noise properties}

\begin{figure}
\center
\includegraphics[width=0.35\textwidth]{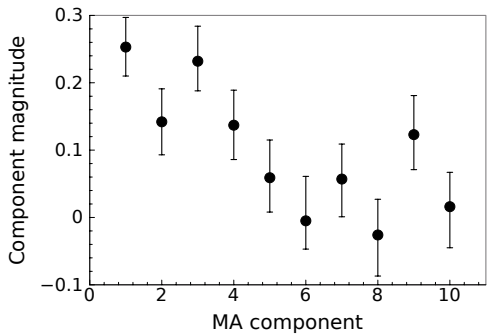}
\caption{Estimated MA components $\phi_{i}$ for the HARPS RVs.}\label{harps_MA}
\end{figure}

The best noise model of the HARPS data was found to be the tenth order MA model with Gaussian white noise component. This model contains 13 free parameters, namely, the magnitude of the Gaussian white noise ($\sigma_{J}$), reference velocity about the data mean $\gamma$, a parameter describing the MA time-scale $\beta$, and ten MA components $\omega_{j}, j=1, ..., 10$. While the MA components $\omega_{j}$ received values between roughly 0.3 and 0.0 indicating that the dependence of the noise of the $i$th measurement on the 10 previous ones was only moderate at most, the time-scale parameter, with units of h$^{-1}$, received an interesting value close to unity with a MAP estimate of 1.19 h$^{-1}$ and a 99\% BCS of [0.98, 1.40] h$^{-1}$. This means that the noise correlations exist on the time-scale of an hour. Also, the MA components roughly decreased as a function of their order in a natural way (Fig. \ref{harps_MA}), which indicates that the correlations between the measurement errors were the greatest the closer these measurements were  in time, both quantitatively and qualitatively. We show the MAP estimates of all the MA parameters in Table \ref{tab:noise_par}.

\begin{table}
\center
\caption{MAP estimates and the corresponding 99\% BCSs of the MA(10) noise parameters.}\label{tab:noise_par}
\begin{tabular}{lc}
\hline \hline
Parameter & HD 10700 b \\
\hline
$\omega_{1}$ & 0.212 [0.168, 0.261] \\
$\omega_{2}$ & 0.120 [0.075, 0.170] \\
$\omega_{3}$ & 0.230 [0.182, 0.282] \\
$\omega_{4}$ & 0.150 [0.097, 0.208] \\
$\omega_{5}$ & 0.071 [0.017, 0.126] \\
$\omega_{6}$ & 0.016 [-0.043, 0.074] \\
$\omega_{7}$ & 0.066 [0.007, 0.126] \\
$\omega_{8}$ & -0.027 [-0.092, 0.031] \\
$\omega_{9}$ & 0.146 [0.080, 0.213] \\
$\omega_{10}$ & 0.010 [-0.054, 0.073] \\
\hline \hline
\end{tabular}
\end{table}

Another interesting parameter, the magnitude of Gaussian white noise, received a low MAP value of 1.06 ms$^{-1}$ (with 99\% BCS of [1.02, 1.11] $^{-1}$) This value is considerably lower than the original 1.7 ms$^{-1}$ we obtained when modelling the noise as pure Gaussian white noise. This indicates that removing the MA components and the three signals from the data reduces the deviation of the data from the mean considerably. It also explains why the three signals can be detected in the data. The reason is that the amplitudes of these signals are only slightly below the noise magnitude enabling the detections, whereas they are considerably below the noise level when the noise correlations are not accounted for. Based on these results and the estimated detection thresholds from the analytical criterion in Eq. (\ref{threshold}), we estimate that with the current precision of HARPS velocities it is already possible to detect signals with RV amplitudes of 0.3-0.5 ms$^{-1}$ if the properties of the noise in the velocities are taken into account.

\subsection{Analysis of partial HARPS data}

To check the robustness of our solution to the HARPS RVs, we tested whether it was possible to receive consistent results with only part of the HARPS velocities.

As our partial data set, we chose the last 2763 HARPS velocities. This choice, while rather arbitrary, was made because it corresponds to excluding the first two observation periods distinguished clearly in Fig. \ref{velocity_fig} (top panel) because of the annual gaps. We decided to exclude the first two observational periods because the overall stability of HARPS has likely been improved after the first two years of operation. We therefore expect, not only that can we extract the same signals from this partial HARPS data, but that we might be able to constrain them better and see the noise levels in the data decrease because of a better stability of the partial data between epochs 3593 and 5082 [JD-2450000].

With this smaller dataset and the same procedures as in the previous section, we identified the same three signals that we found in the full HARPS data set. We could not find a clear probability maximum for a four-Keplerian model. This confirms that there are three statistically significant signals in the data. When we increased the number of Keplerian signals in the statistical model from zero to three, the posterior probabilities increased by factors of 3.9$\times10^{16}$, 3.5$\times10^{10}$, and 3.4$\times10^{9}$, which indicates that these signals are again very significant. However, comparing these values to the factors received for the full data set indicates that, while they are very similar when moving from $k=0$ to $k=2$, the significance of the third signal decreases considerably for the partial HARPS data.

These results are consistent with the ones received for the full HARPS data set but we found a significantly different noise level in the partial HARPS data set. The parameter $\sigma_{J}$ that describes the magnitude, i.e. standard deviation, of the Gaussian excess white noise in the data, received a MAP estimate of 0.82 ms$^{-1}$ (with a 99\% BCS of [0.78, 0.87] ms$^{-1}$). This can be compared to the MAP estimate of 1.05 ms$^{-1}$ for the full HARPS data set (Table \ref{3Kep_HARPS}). Thus, the first two observing periods indeed contain considerably more noise than the subsequent ones. This could be due to improvements in the instrumental stability of the HARPS spectrograph over the years, but we cannot know this for sure as it might also be caused by a more quiescent period of the target star with e.g. a lower amount of starspots. Either way, it looks like the first two observing periods are contaminated by a significantly greater amount of noise than the subsequent periods.

\subsection{HARPS activity indicators}\label{sec:activity}

Our Bayesian analyses identify three clear periodic signals -- possibly caused by periodic Doppler shifts -- in the RV data acquired using HARPS, but the possibility of these signals being caused by activity-related phenomena on the stellar surface must be accounted for. Therefore, we extracted the HARPS $S$ chromospheric activity indices using methods honed using other high resolution spectrographs \citep{jenkins2006,jenkins2011,tuomi2012c} to search for periodicities in the activities, and/or correlations between the activity of HD10700 and the RV signals we find.

\begin{figure}
\center
\includegraphics[angle=90,width=0.35\textwidth]{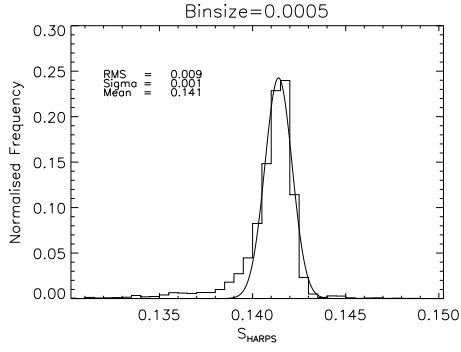}
\caption{The distribution of HARPS $S$-index. The solid curve indicates the fitted Gaussian curve.}\label{act_dis}
\end{figure}

The distribution of HARPS $S$-indices for HD 10700 was found to have an approximately Gaussian profile, as shown in Fig. \ref{act_dis}. We can see that HD 10700 has a very tight spread of chromospheric activities which helps us to realise that this star is magnetically very stable. The standard deviation of the $S$-index distribution (Fig. \ref{act_dis}) is only 0.001 dex, which is very stable in comparison to other typical G-dwarf stars. For instance, the Sun can exhibit activity index changes at the level of 0.005 dex over long timescales \citep{livingston2007}. This suggests that HD 10700 is exhibiting some period of sustained magnetic stability, or its orientation in space is such that it appears from our vantage point that the spot patterns on the stellar surface are of very low number and do not change considerably over the baseline of the HARPS measurements.

Another feature worth noting is that the distribution of the $S$-index is well described by a Gaussian function except in the lower wing region. There appears to be a small excess of low activity values for HD 10700 that may require an additional modification of the best fit distribution. It may be that a double Gaussian is needed, similar to the distribution seen for HD 114613 \citep{tuomi2012c} but at a much more reduced level. This may indicate that double Gaussian distributions of $S$-indices are common for F, G, and K dwarf and subgiant stars over these timescales of few years at these levels of precision.

\begin{figure}
\center
\includegraphics[angle=90,width=0.35\textwidth]{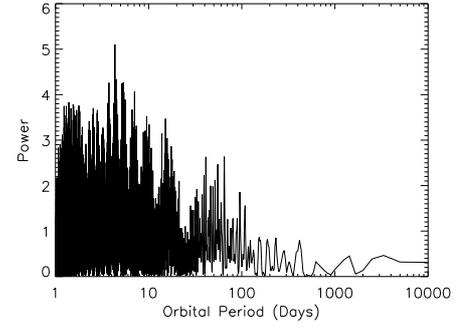}

\includegraphics[angle=90,width=0.35\textwidth]{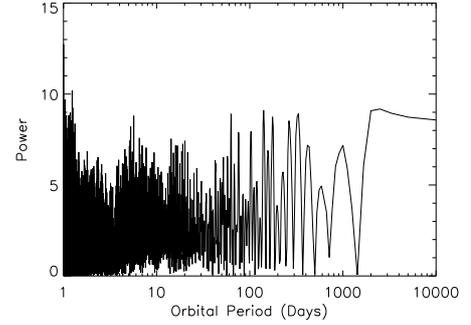}
\caption{Lomb-Scargle periodograms of the $S$-indices (top panel) and the BIS values (bottom panel).}\label{activity_periodogram}
\end{figure}

The top panel in Fig. \ref{activity_periodogram} shows a Lomb-Scargle periodogram analysis of the $S$-indices and there only appears to be one region in the period space that exhibits strong signals compared to the rest of the data. The strongest power in this periodogram has a period of 4.34 days and the eight strongest powers are found in a range between 3.7 and 5.5 days. Interestingly, we detected a weak signal, i.e. a clear probability maximum but not a statistically significant periodicity, in the partial HARPS data with a period of 3.70 days. Taking into account the features in the activity data, we expect that this short period signal is likely caused by activity-related features in the RV data and not by a genuine Doppler shift of planetary origin.

The bottom panel in Fig. \ref{activity_periodogram} shows the periodogram for the bisector span (BIS) values of the HARPS data set. These BIS values were drawn directly from the HARPS-DRS analysis and details of their usefulness can be found in \citet{santos2010}. There are no significant periodicities in the BIS values data either. These results, i.e. the lack of significant periodicities in both, the $S$ indices and BIS values, indicate that activity or line asymmetries are not the cause of the periodicities we detect in the HARPS RVs. Nevertheless, we should note that our 35.362 day signal is close to the 34 day period (Table \ref{properties}) reported by \citet{baliunas1996} and is something we discuss further in Section \ref{sec:discussion}.

Since the signals we have discovered in the RVs are of very low amplitude we can investigate whether activity indicators, i.e. the $S$-index and the BIS value, correlate with the RV variations in the HD10700 data. We calculated the phase-folded signals (as shown in Fig. \ref{orbits}) and searched for such correlations between each of the signals and the activity indicators. We could not find any strong linear correlations between these indices and any of the signals. None of these correlations were found to be significant and the corresponding Pearson correlation coefficients do not exceed $\pm$0.08. This supports the argument that spot modulation, or some other periodic stellar phenomena that affects the stellar line profiles, is not the root cause of the periodicities that we find in the data for HD 10700, reinforcing the possibility that the three signals might be induced by planets orbiting the star.

\section{AAPS and HIRES radial velocities}\label{sec:other_analyses}

The AAPS and HIRES velocities had significantly more noise -- roughly three times more -- than the HARPS precision velocities, which suggests that detecting the same signals might not be possible from them. We started by finding the best noise model for the AAPS and HIRES RVs. We compared the performance of different MA models because the AR models could not be considered trustworthy descriptions of RVs based on the tests with artificially injected signals. According to our results, the fifth order MA model was the best description of the AAPS data and increasing the number of MA components did not result in a model with a greater posterior probability. For HIRES data, the first order MA model was already a reasonably accurate description and increasing the order above three resulted in only a minor improvement that we do not consider significant. The posterior probabilities of the statistical models are shown in Table \ref{AH_probs} up to a fifth order MA model. Based on these posterior probabilities, we use the MA(5) and MA(3) models for the AAPS and HIRES data in the following analyses.

\begin{table}
\center
\caption{Relative posterior probabilities of noise models for AAPS and HIRES RVs.}\label{AH_probs}
\begin{tabular}{lcc}
\hline \hline
Model & $P(\mathcal{M} | d)$ (AAPS) & $P(\mathcal{M} | d)$ (HIRES) \\
\hline
G & 7.8$\times10^{-104}$ & 4.5$\times10^{-28}$ \\
G+MA(1) & 2.7$\times10^{-28}$ & 0.04 \\
G+MA(3) & 9.8$\times10^{-5}$ & 0.24 \\
G+MA(5) & $\sim$ 1 & 0.72 \\
\hline \hline
\end{tabular}
\end{table}

\begin{figure}
\center
\includegraphics[angle=-90,width=0.45\textwidth,clip]{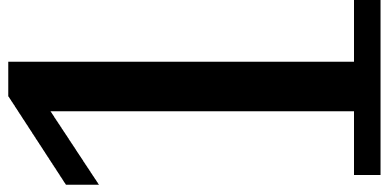}

\includegraphics[angle=-90,width=0.45\textwidth,clip]{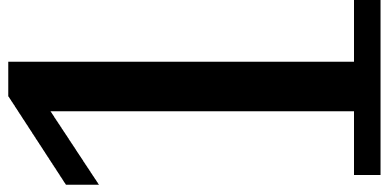}
\caption{Lomb-Scargle periodograms of the AAPS (top) and HIRES (bottom) data sets after removing the noise correlations from them. Dotted, dashed, and dash-dotted lines indicate the analytical 10\%, 1\%, and 0.1\% FAPs, respectively.}\label{AH_periods}
\end{figure}

After removal of the identified noise components, we could neither identify any significant periodic signals in the AAPS and HIRES RVs nor significant powers in their periodograms (Fig. \ref{AH_periods}). Despite several samplings of the parameter space of a one-Keplerian model, our Markov chains did not converge to any periodicities for either data set. This indicates, that the signals detected from the HARPS velocities are below the detection thresholds of these two data sets and cannot be obtained from them.

The nature of the noise in the AAPS and HIRES data sets turned out to be different from that observed in the HARPS RV data. The noise in AAPS and HIRES data sets did not require as many MA components as HARPS data did and the time-scale of the correlations turned out to be different as well. We find that the $\alpha$ parameter in Eq. (\ref{correlation_function}) received values of 0.0026 [0.0005, 0.0080] h$^{-1}$ and 0.0244 [0.0006, 0.0740] h$^{-1}$ for the AAPS and HIRES velocities, respectively. These values correspond to noise correlations on the time-scale of days to weeks, not the few hour time-scale of the HARPS data.

This different time-scale of noise correlations in the AAPS and HIRES data sets could be caused by e.g. instability of the instruments from night-to-night that exceeds any noise correlations on shorter time-scales. Indeed asterioseismology analyses suggest that the short-term performance of the spectrographs are relatively similar \citep{bedding2007}. The HARPS spectrograph is shown to have good short-term stability \citep[the reader is referred to the long series of publications titled ''The HARPS search for southern extra-solar planets`` of which two recent ones are:][]{dumusque2011,segransan2011}, which could enable the detection of noise correlations over time-scales of a few hours -- possibly arising from stellar surface phenomena -- from HARPS data.

Whether caused by the telescope-instrument combinations or the surface of the stellar target, the noise correlations need to be taken into account when analysing AAPS and HIRES RVs. While their inclusion in the statistical model improves the performance of the model considerably (Table \ref{AH_probs}), it also enables us to remove these correlations from the data, which could reveal signals otherwise hidden in the noise. For a pure Gaussian noise model, the magnitude of the excess jitter was found to be 3.18 [2.90, 3.48] ms$^{-1}$ and 2.45 [2.17, 2.74] ms$^{-1}$ for the AAPS and HIRES data sets, respectively. These values decreased to 2.16 [1.93, 2.44] ms$^{-1}$ and 2.11 [1.86, 2.42] ms$^{-1}$ for the best MA models, which indicates that a considerable amount of noise that results from correlations was interpreted as Gaussian white noise in the pure Gaussian white noise model. Therefore, as was the case for the HARPS RV data, accurate noise modelling can improve the potential of the AAPS and HIRES programmes to find lower amplitude signals.

We note that the white noise component appeared to be close to Gaussian in the AAPS and HIRES data sets. The Cauchy model provided a much worse description of these data sets because they lack considerable outliers. Also, the noise model in Eq. \ref{convolution} was worse than the pure Gaussian model because it has one extra parameter that appears to be unnecessary because the noise distribution does not have a flat maximum.

\section{Combined radial velocities}\label{sec:combined_analyses}

We combined the HARPS, AAPS, and HIRES RV data in an attempt to see if the signals detected in the HARPS data alone could be extracted from this combined set as well, and especially, whether their significances would change with respect to those found in the HARPS data alone. For obvious reasons, the combined data set had a better phase coverage than any of the individual data sets and a baseline corresponding to that of the AAPS data of approximately 13.5 years. We modelled the noise properties of this combined set by using the best noise descriptions found for each data set. While the periodogram of the combined data did not show any signatures of periodic signals after removing the correlations in the noise, we performed samplings of models with $k$ Keplerian signals, with $k=1, 2, ...$, anyway. 

Identifying the 35 day periodicity was again straightforward and the corresponding one-Keplerian model received a posterior probability that was 8.1$\times 10^{13}$ times greater than that of the model without Keplerian signals (see Table \ref{combined_probabilities}). Similarly, adding more Keplerian signals in the statistical model we could identify the 13.9 and 94 day periodicities rather easily. Taking these periodicities into account increased the model probabilities further by factors of 2.6$\times 10^{11}$ and 2.5$\times 10^{15}$, respectively, which means that the three Keplerian signals detected from the HARPS data alone were present in the combined data set as well with high significances. However, we note that with the three-Keplerian model, these three signals did not receive exactly the same parameter estimates as they did for the HARPS data alone. Their RV amplitudes received values of approximately 0.10 ms$^{-1}$ lower than for the HARPS data and their eccentricities received slightly lower values better consistent with circular orbits. Yet, given their 99\% BCSs, all parameter values were consistent with the estimates in Table \ref{3Kep_HARPS}. We note that these observed differences in the MAP estimates from HARPS data and the combined data could be caused by insufficient noise modelling.

\begin{figure}
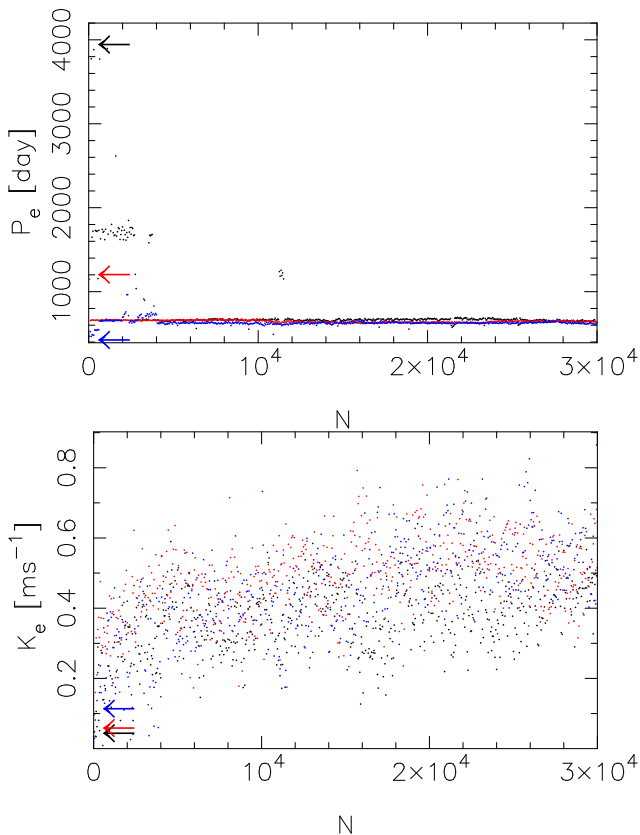

\center
\includegraphics[angle=-90,width=0.45\textwidth]{rvdist04_rv_HD10700_multiw_Pe.ps}
\includegraphics[angle=-90,width=0.45\textwidth]{rvdist04_rv_HD10700_multiw_Ke.ps}
\caption{Convergence of the parameters of the fourth Keplerian as a function of Markov chain length to a period of 630 days (top) and a RV amplitude of 0.56 ms$^{-1}$ (bottom) for three different Markov chains (denoted using different colours). Arrows indicate the randomly selected initial values ($K$ is chosen to be initially close to zero). The chains have been thinned by a factor of ten.}\label{4p_convergence}
\end{figure}

We continued by sampling the parameter space of a four-Keplerian model, and despite the fact that none of the residual periodograms of the individual data sets or the full set showed any powers in addition to the three aforementioned signals, we discovered a significant probability maximum in the parameter space corresponding to another signal with a period of 630 days and a RV amplitude of 0.52 [0.24, 0.85] ms$^{-1}$. This signal satisfied all our detection criteria by making the four-Keplerian model 1.3$\times 10^{5}$ times more probable than the three-Keplerian one. Also, all the MCMC samplings we performed with different initial states converged to this same solution (Fig. \ref{4p_convergence}) and we could safely conclude that the 630 days periodicity corresponds to a reasonably high probability maximum and is one of the periods in the four-Keplerian model. We note that this 630 day signal shows in the periodogram of the HARPS data as well before removing any periodicities from the data as a peak exceeding the 10\% FAP (Fig. \ref{periodograms}, top panel).

\begin{table*}
\center
\caption{Relative posterior probabilities and log-Bayesian evidences of models $\mathcal{M}_{k}$ with $k$ Keplerian signals given the combined HARPS, AAPS, and HIRES RV data. $P_{s}$ denotes the MAP period estimate of the signal added to the solution.\label{combined_probabilities}}
\begin{tabular}{lccccc}
\hline \hline
$k$ & $P(\mathcal{M}_{k} | d)$ & $\ln P(d | \mathcal{M}_{k})$ & $P(\mathcal{M}_{k} | d)$ & $\ln P(d | \mathcal{M}_{k})$ & $P_{s}$ [days] \\
 & TPM & TPM & AICc & AICc & \\
\hline
0 & 5.0$\times10^{-53}$ & -10689.4 & 6.5$\times10^{-49}$ & -10699.0 & -- \\
1 & 4.0$\times10^{-39}$ & -10656.7 & 2.3$\times10^{-36}$ & -10669.4 & 35.2 \\
2 & 1.1$\times10^{-27}$ & -10629.7 & 2.3$\times10^{-26}$ & -10645.6 & 13.9 \\
3 & 2.6$\times10^{-12}$ & -10593.6 & 1.4$\times10^{-11}$ & -10610.9 & 95.4 \\
4 & 3.3$\times10^{-7}$ & -10581.1 & 8.3$\times10^{-6}$ & -10596.9 & 630 \\
5 & 0.937 & -10565.6 & $\sim$ 1 & -10584.5 & 168 \\
5 & 0.063 & -10568.3 & 3.1$\times10^{-4}$ & -10592.6 & 315 \\
\hline \hline
\end{tabular}
\end{table*}

When sampling the parameter space of a five-Keplerian model, we chose the four-Keplerian one as a starting point and set the initial parameter values corresponding to the four signals close to their MAP estimates. However, the period of the fifth one was chosen randomly either between the 95 and 640 days or above 630, such that its velocity amplitude was set close to zero. This division of the period-space into two parts was made because we could then treat the corresponding models with the fifth periodicity in either subspace as separate models. We chose this division because our initial samplings indicated that both subspaces contained probability maxima and sampling a parameter space with well-separated modes is computaionally very demanding. The division enabled us to perform the corresponding samplings much more rapidly than performing the samplings of the full period space.

Samplings of the parameter space identified three additional probability maxima in the period space at roughly 170, 320, and 1300 days. These periods correspond to orbital distances where low-mass planets would likely remain on stable orbits because they retain a sufficient orbital spacing. Also, we note that the 320 day signal shows as a peak exceeding the 10\% FAP in Fig. \ref{periodograms} (top panel) but because of the fact that this peak is so close to one year periodicity, which also shows in the window function as a result of HARPS data sampling (Fig. \ref{HARPS_periodogram}, bottom panel), we cannot conclude that it actually corresponds to a genuine signal.

According to our samplings of the parameter space of the five-Keplerian model, the probability maximum around 1300 days did not turn out to be significant. The amplitude of the signal corresponding to this probability maximum was not found distinguishable from zero, and we could conclude that the data, when interpreted using the five-Keplerian model, did not support the existence of a signal at or around 1300 days. Also, though there are other lower probability maxima beyond 1300 day period (especially around 2000 days) up to the maximum period in the parameter space of ten times the data baseline, none of them was found significant either.

\begin{figure}
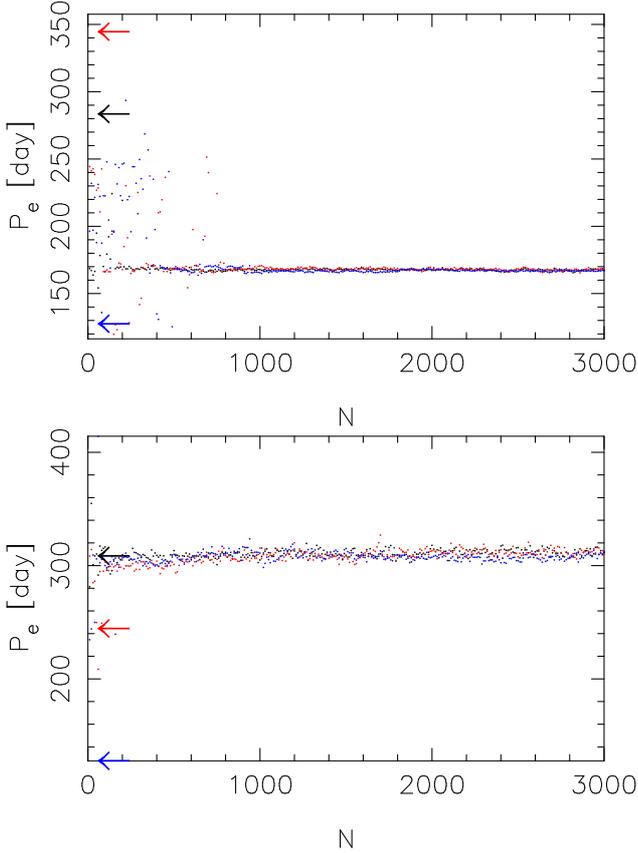

\center
\includegraphics[angle=-90,width=0.45\textwidth]{rvdist05_rv_HD10700_multiw_Pe_170.ps}
\includegraphics[angle=-90,width=0.45\textwidth]{rvdist05_rv_HD10700_multiw_Pe_320.ps}
\caption{Convergence of the period of the fifth Keplerian as a function of Markov chain length to periods of 168 (top) and 315 days (bottom). The details of the panels are as in Fig. \ref{4p_convergence} but the chains have been thinned by a factor of ten.}\label{5p_period_convergence}
\end{figure}

Instead, the period space between 95 and 630 days provided some interesting solutions for the fifth signal. Especially, we found a global probability maximum for the fifth period at 168 days which increased the model probability of the five-Keplerian model a factor of $3.0 \times 10^{6}$ greater than that of the four-Keplerian one (Table \ref{combined_probabilities}). In addition to this solution, we identified another possible solution for the fifth signal at 315 days. This local solution was found to have a significantly greater probability than the four-Keplerian model did, but it was not as probable as the global solution corresponding to the 168 day periodicity as a fifth signal (Table \ref{combined_probabilities}). Yet, both these solutions satisfied all the detection criteria by having RV amplitudes statistically significantly greater than zero and by having well-constrained periods and the Markov chains converged to either one of them very rapidly regardless of the initial choice of the fifth period (Fig. \ref{5p_period_convergence}). We note that 168 and 315 day periodicities are actually each other's one-year aliases. We could easily verify this by sampling a six-Keplerian model with these two periodicities as initial states of the periods of two signals. These samplings did not converge to two signals but altered between either of the periodicities in the sense that the RV amplitudes of the 168 and 315 day signals had a negative correlation with either reaching a maximum when the other approached zero.

The five-Keplerian solution with the 168 day periodicity is clearly favoured by the data according to the model probabilities in Table \ref{combined_probabilities}. However, the solution with the 315 day periodicity still has a considerable probability of 0.06 (though it is not significant based on the AICc), which means that we cannot completely rule out this alternative solution. Because of this alternative solution and a local maximum in the period space of the fifth Keplerian signal at roughly 1300 days, we carried on sampling the parameter space of a six-Keplerian model further.

The global solution of the six-Keplerian model was found to correspond to periodicities at 14, 35, 94, 168, 640, and 1300 days. However, despite the fact that the global solution contained the 1300 day signal, this signal was found to have an amplitude that was not statistically significantly different from zero. Also, We could not identify a solution where the sixth signal had a period between the 168 and 640 day ones. In the six-Keplerian solution, the periodicity of 1300 days was well constrained from above and below, but the RV amplitude of this signal was only barely constrained with a MAP estimate of 0.36 ms$^{-1}$ and a 99\% BCS of [0.00, 0.66] ms$^{-1}$, which demonstrates that there is a fair chance that this signal in fact has a negligible amplitude, i.e. there is no evidence in favour of its existence. Because we could not distinguish this signal in the six-Keplerian model from zero, we did not calculate the probability for this model because we could not be sure whether the corresponding Markov chains had converged.

\begin{figure*}
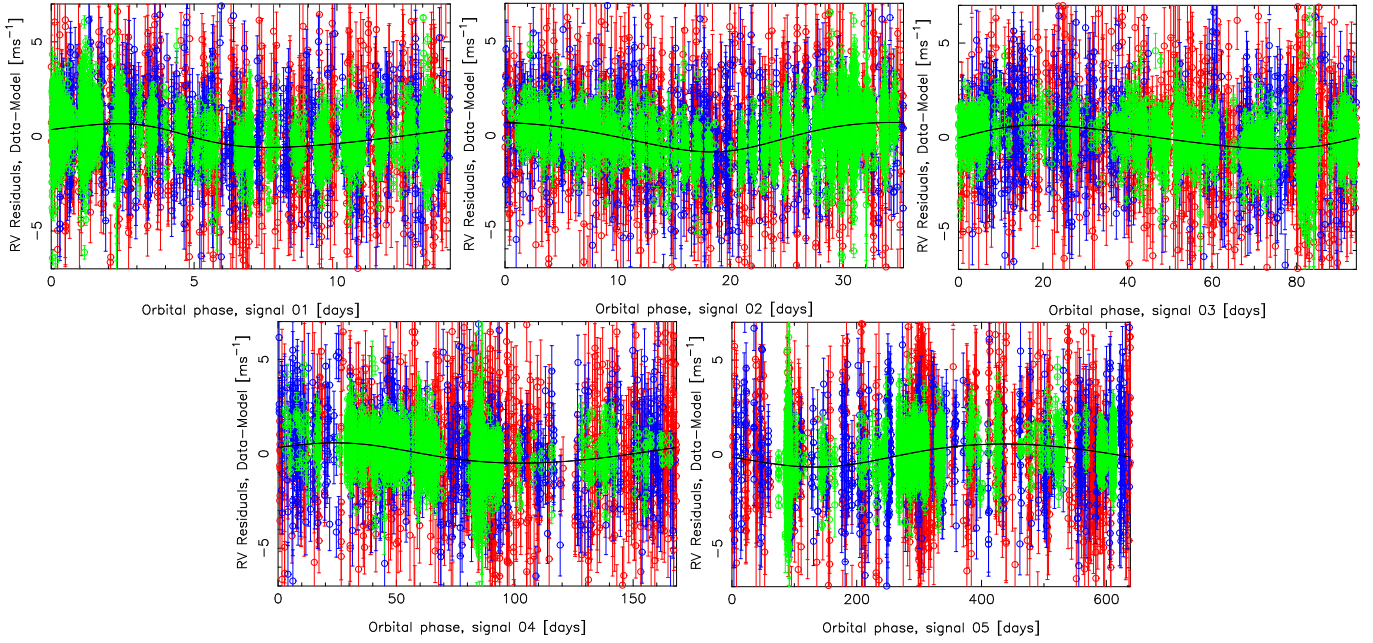

\center
\includegraphics[angle=-90,width=0.32\textwidth]{rvdist05_scresidc_rv_HD10700_1.ps}
\includegraphics[angle=-90,width=0.32\textwidth]{rvdist05_scresidc_rv_HD10700_2.ps}
\includegraphics[angle=-90,width=0.32\textwidth]{rvdist05_scresidc_rv_HD10700_3.ps}

\includegraphics[angle=-90,width=0.32\textwidth]{rvdist05_scresidc_rv_HD10700_4.ps}
\includegraphics[angle=-90,width=0.32\textwidth]{rvdist05_scresidc_rv_HD10700_5.ps}
\caption{Phase-folded Keplerian signals in the combined HARPS (green), AAPS (red), and HIRES (blue) RVs of HD 10700. Vertical axis is scaled to the interval of [-7, 7] ms$^{-1}$ for clarity and the RVs deviating more than that from zero are therefore not shown.}\label{combined_orbits}
\end{figure*}

\begin{figure*}
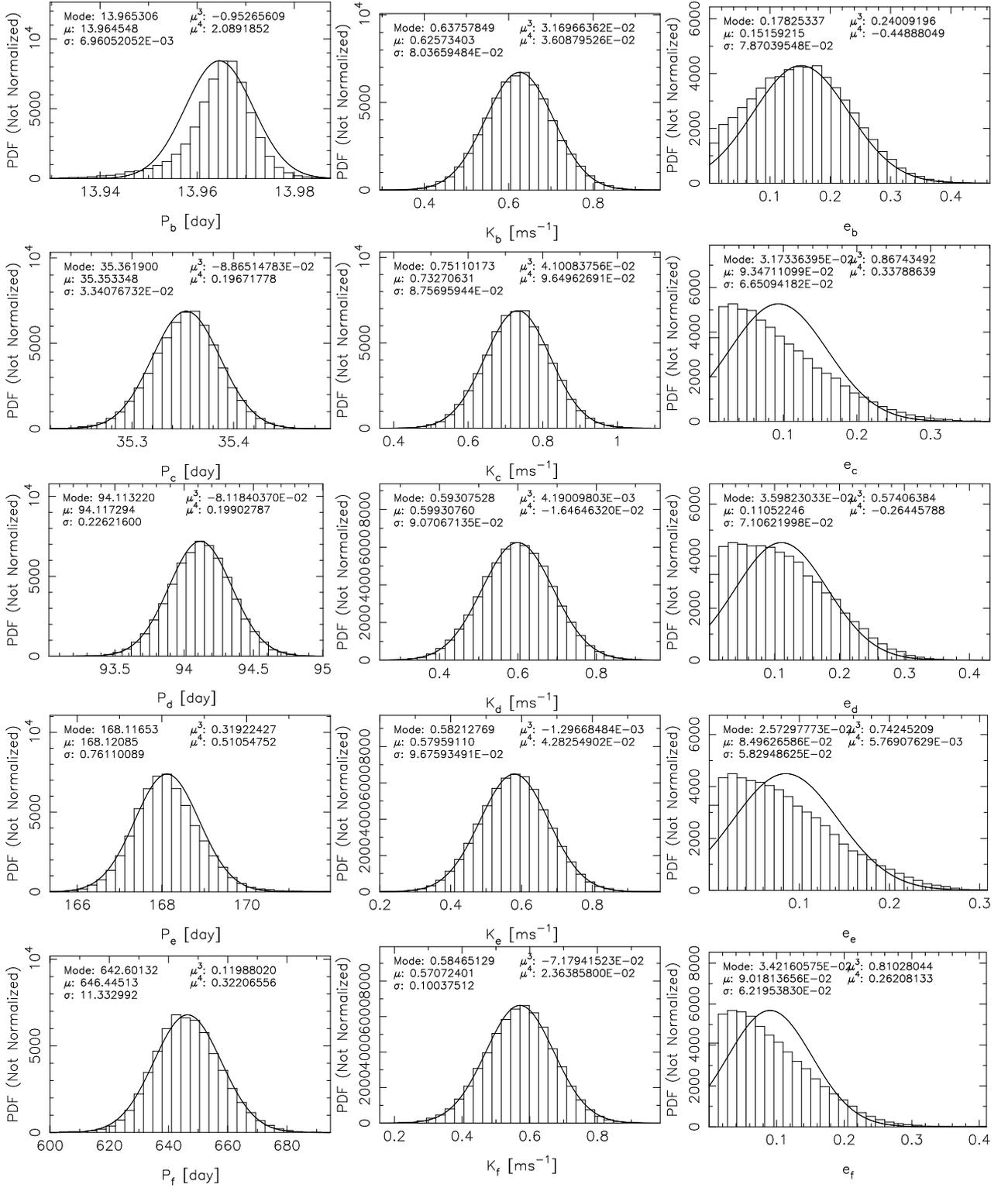

\center
\includegraphics[angle=-90,width=0.3\textwidth]{rvdist05_rv_HD10700_dist_Pb.ps}
\includegraphics[angle=-90,width=0.3\textwidth]{rvdist05_rv_HD10700_dist_Kb.ps}
\includegraphics[angle=-90,width=0.3\textwidth]{rvdist05_rv_HD10700_dist_eb.ps}

\includegraphics[angle=-90,width=0.3\textwidth]{rvdist05_rv_HD10700_dist_Pc.ps}
\includegraphics[angle=-90,width=0.3\textwidth]{rvdist05_rv_HD10700_dist_Kc.ps}
\includegraphics[angle=-90,width=0.3\textwidth]{rvdist05_rv_HD10700_dist_ec.ps}

\includegraphics[angle=-90,width=0.3\textwidth]{rvdist05_rv_HD10700_dist_Pd.ps}
\includegraphics[angle=-90,width=0.3\textwidth]{rvdist05_rv_HD10700_dist_Kd.ps}
\includegraphics[angle=-90,width=0.3\textwidth]{rvdist05_rv_HD10700_dist_ed.ps}

\includegraphics[angle=-90,width=0.3\textwidth]{rvdist05_rv_HD10700_dist_Pe.ps}
\includegraphics[angle=-90,width=0.3\textwidth]{rvdist05_rv_HD10700_dist_Ke.ps}
\includegraphics[angle=-90,width=0.3\textwidth]{rvdist05_rv_HD10700_dist_ee.ps}

\includegraphics[angle=-90,width=0.3\textwidth]{rvdist05_rv_HD10700_dist_Pf.ps}
\includegraphics[angle=-90,width=0.3\textwidth]{rvdist05_rv_HD10700_dist_Kf.ps}
\includegraphics[angle=-90,width=0.3\textwidth]{rvdist05_rv_HD10700_dist_ef.ps}
\caption{Distributions estimating the posterior densities of orbital periods ($P_{x}$), eccentricities ($e_{x}$), and RV amplitudes ($K_{x}$) of the five Keplerian signals. The solid curve is a Gaussian density with the same mean ($\mu$) and variance ($\sigma^{2}$) as the parameter distribution has. Additional statistics, mode, skewness ($\mu^{3}$) and kurtosis ($\mu^{4}$) of the distributions are also shown.}\label{densities}
\end{figure*}

To make sure that we did not miss any additional signals, we performed samplings of the parameter space of a seven-Keplerian model. These samplings did not help identifying any other significant probability maxima in the parameter space. Therefore, we conclude that the best description of the combined velocities of HD 10700 contains five Keplerian signals. We present this orbital solution in Table \ref{5Kep_combined} and show the phase-folded signals in Fig. \ref{combined_orbits}. We also show the distributions of orbital periods, RV amplitudes, and eccentricities in Fig. \ref{densities} to demonstrate that the periods and amplitudes are well constrained and, especially, fully comply with our detection criteria. In order to visualise the significance of the signals, we plotted the phase-folded orbits by dividing the phase of each signal into 200 bins and by calculating the means and standard deviations for each bin. These plots are shown in the Appendix for the combined data set and for the HARPS data alone (Figs. \ref{bin_orbits_all} and \ref{bin_orbits_harps}). When comparing these two figures, it can also be seen how the phase coverages of the signals are improved when using the combined data set instead of the HARPS data alone.

\begin{table*}
\center
\caption{Five-planet solution of HD 10700 RVs. MAP estimates of the parameters and their 99\% BCSs. Parameter $t_{0}$ denotes the time of periastron.\label{5Kep_combined}}
\begin{tabular}{lccc}
\hline \hline
Parameter & HD 10700 b & HD 10700 c & HD 10700 d \\
\hline
$P$ [days] & 13.965 [13.941, 13.982] & 35.362 [35.256, 35.450] & 94.11 [93.48, 94.81] \\
$e$ & 0.16 [0, 0.38] & 0.03 [0, 0.31] & 0.08 [0, 0.34] \\
$K$ [ms$^{-1}$] & 0.64 [0.40, 0.88] & 0.75 [0.48, 0.99] & 0.59 [0.34, 0.87] \\
$\omega$ [rad] & 1.5 [0, 2$\pi$] & 3.0 [0, 2$\pi$] & 4.0 [0, 2$\pi$] \\
$M_{0}$ [rad] & 2.6 [0, 2$\pi$] & 3.2 [0, 2$\pi$] & 5.8 [0, 2$\pi$] \\
$t_{0}$ [days]$^{\star}$ & 4.17 [-] & 20.62 [-] & 2.31 [-] \\
$m_{p} \sin i$ [M$_{\oplus}$] & 2.0 [1.2, 2.8] & 3.1 [2.0, 4.5] & 3.6 [1.9, 5.3] \\
$a$ [AU] & 0.105 [0.099, 0.110] & 0.195 [0.184, 0.204] & 0.374 [0.354, 0.391] \\
\hline
& HD 10700 e & HD 10700 f \\
\hline
$P$ [days] & 168.12 [166.01, 170.44] & 642 [615, 679] \\
$e$ & 0.05 [0, 0.27] & 0.03 [0, 0.29] \\
$K$ [ms$^{-1}$] & 0.58 [0.30, 0.86] & 0.58 [0.26, 0.85] \\
$\omega$ [rad] & 5.5 [0, 2$\pi$] & 3.9 [0, 2$\pi$] \\
$M_{0}$ [rad]& 0.5 [0, 2$\pi$] & 1.6 [0, 2$\pi$] \\
$t_{0}$ [days]$^{\star}$ & 37.42 [-] & 168.49 [-] \\
$m_{p} \sin i$ [M$_{\oplus}$] & 4.3 [2.2, 6.3] & 6.6 [3.1, 10.1] \\
$a$ [AU] & 0.552 [0.522, 0.575] & 1.35 [1.26, 1.43] \\
\hline
$\sigma_{J,1}$ [ms$^{-1}$] (HIRES) & 2.14 [1.92, 2.41] \\
$\sigma_{J,2}$ [ms$^{-1}$] (AAPS) & 2.13 [1.88, 2.41] \\
$\sigma_{J,3}$ [ms$^{-1}$] (HARPS) & 1.06 [1.02, 1.10] \\
\hline \hline
\end{tabular}
\tablefoot{($\star$) Days after JD$=$2450000 that was used as the zero-point in our analyses.}
\end{table*}

The five signals we observe in the combined data of HD 10700 are all consistent with circular orbits and have MAP estimated amplitudes below 1 ms$^{-1}$. Despite these low amplitudes, their detection is robust in the sense that (1) their inclusion in the statistical model increases the model probabilities significantly, (2) they have well-constrained periods and amplitudes that differ statistically from zero. We have shown in Section \ref{sec:activity} that these signals do not have counterparts in the S and BIS activity indicators. It also seems unlikely that they arise from data sampling since the phase-folded RV curves have a good phase-coverage (Fig. \ref{combined_orbits}) and the signals are constrained reasonably accurately (Fig. \ref{densities} and Table \ref{5Kep_combined}).

\subsection{Signals in periodograms}

\begin{figure*}
\center
\includegraphics[angle=-90,width=0.49\textwidth]{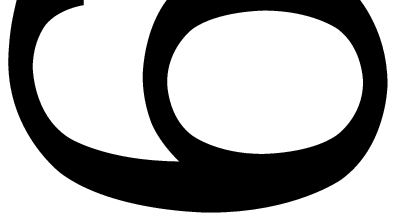}
\includegraphics[angle=-90,width=0.49\textwidth]{tauceti_5p_3.ps}

\includegraphics[angle=-90,width=0.49\textwidth]{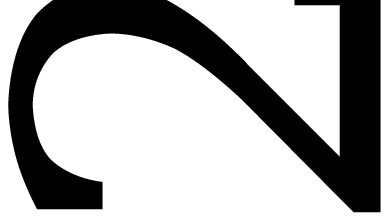}
\includegraphics[angle=-90,width=0.49\textwidth]{tauceti_5p_4.ps}

\includegraphics[angle=-90,width=0.49\textwidth]{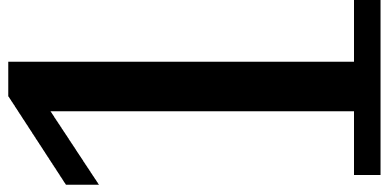}
\includegraphics[angle=-90,width=0.49\textwidth]{tauceti_5p_5.ps}
\caption{Lomb-Scargle periodograms of the combined data residuals after removing MA components from the noise (top left panel) and removing the 35 (middle left panel), 14 (bottom left panel), 95 (top right panel), 168 (middle right panel), and 640 day signals (right bottom panel). The dotted, dashed, and dot-dashed lines indicate the analytical 10\%, 1\%, and 0.1\% FAPs, respectively.}\label{combined_periodograms}
\end{figure*}

To examine whether the three most significant signals are indeed present in the data and whether the two longer periodicities of 168 and 640 days are distinguishable from the data, we subtracted the MA components of the best model in our analyses from the combined data and analysed the residuals using the Lomb-Scargle periodograms. The sequence of residual periodograms is shown in Fig. \ref{combined_periodograms}.

In Fig \ref{combined_periodograms}, the top left panel shows the residual periodogram after removing the MA components from the data. This periodogram shows several features exceeding the 0.1\% FAP level. The highest peak corresponds to a periodicity at 35 days and has an analytical FAP of 2.2$\times 10^{-69}$. Similarly, after removing the most significant periodicities and again calculating the residual periodograms the one- and two-signal residuals show power at 13.9 and 95 days (left middle and bottom panels) exceeding the 1.5$\times 10^{-27}$ and 1.4$\times 10^{-19}$ FAPs, respectively. These three signals correspond to the ones detected from the HARPS data alone.

In Fig. \ref{combined_periodograms}(right top panel) we calculate the residual periodogram of the three-signal model.This periodogram has two significant powers exceeding the 1\% FAP level at 15 and 168 days. Because we did not find signals at 15 days in our analyses, we adopted the 168 day periodicity as a fourth signal and carried on by calculating the residual periodogram of the four-signal model (Fig. \ref{combined_periodograms}, right top panel). Removing the 168 day signal also removed the peak at 15 days, which indicates that the latter was an alias of the former. The periodogram shows the 640 day periodicity as a last significant power (with a FAP of 2.5$\times 10^{-3}$, Fig. \ref{combined_periodograms} right middle panel). Based on these results, the periodogram analyses support the results of the Bayesian analyses that there are five periodic signals in the combined RV data of HD 10700 on the basis that the noise in the data is modelled by using the MA models and Gaussian white noise.

\section{Dynamical stability of the Keplerian solution}\label{sec:dynamics}

\begin{figure}
\center
\includegraphics[angle=-90,width=0.45\textwidth]{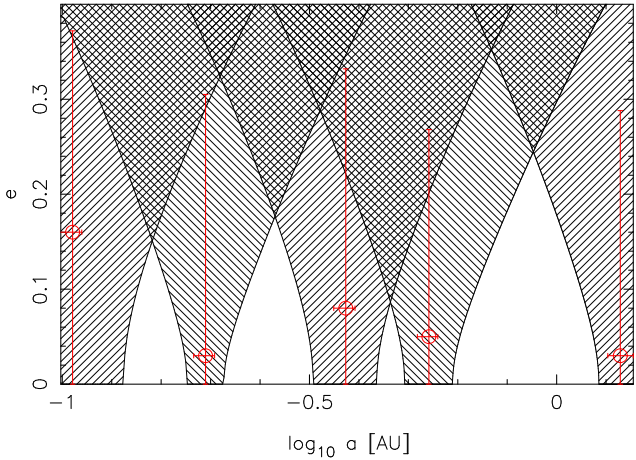}
\caption{Approximate Lagrange stability thresholds calculated for the HD 10700 signals. The hatched areas surrounding the orbital MAP estimates (red circles) indicate the parameter space where the neighbouring planets would make the system unstable.}\label{fig:stability}
\end{figure}

The discovery of these periodic signals has only been possible as a result of noise modelling and indeed we can not be confident of this procedure because we cannot find the signals independently in the different datasets. We nonetheless consider it instructive to assess whether these periodic signals correspond to a planetary system that is dynamically stable in long-term. Thus, following \citet{tuomi2012}, we investigated the stability of the planetary system corresponding to the five signals by plotting the approximate Lagrange stability thresholds for each subsequent pair of planets in the system \citep{barnes2006}. According to our results shown in Fig. \ref{fig:stability}, the system appears to be long-term stable if the orbital eccentricities are not much greater than their MAP estimates (see Table \ref{5Kep_combined}). Fig. \ref{fig:stability} also indicates that there are orbital distances where additional low-mass planets could remain on stable orbits.

We tested the predictions of the stability thresholds by assuming that the putative planets in the HD 10700 system have co-planar orbits. With this assumption, we increased the masses corresponding to the inclination of the orbital plane approaching zero. According to our results, if the eccentricities of the five planets are at or below their MAP estimates, the stability thresholds are still in favour of a dynamically stable system when the inclination of the orbital plane is below 3$^{\circ}$. This corresponds to actual masses of roughly 40 times the minimum masses reported in Table \ref{5Kep_combined}. Therefore, the orbital spacing of the putative five-planet system around HD 10700 enables almost any orbital inclinations, and thus the system can still be dynamically stable on long timescales for masses considerably greater than the minimum masses. We note that full-scale numerical integrations of the orbits of the putative planets are needed to verify the above results and are beyond the scope of this paper.

\section{Discussion}\label{sec:discussion}

Signals with amplitudes lower than 1 ms$^{-1}$ have been reported in quiescent HARPS targets such as HD 20794 and HD 85512 \citep{pepe2011}. The lowest reported RV amplitude in the literature appears to be the 0.56 ms$^{-1}$ one of HD 20794 c \citep{pepe2011}. In practice, such accuracy corresponds to being able to find habitable zone super-Earths and hot rocky planets even less massive than the Earth orbiting nearby Solar-type stars.

In the current work, we have considered several noise models for high-precision velocities using a large data set for HD10700 from the HARPS, AAPS and Keck exoplanet surveys. Our tests indicate that noise models containing a moving average component and assuming Gaussian white noise is the most effective modelling strategy for such data. Our simulations of the HARPS data for HD10700 indicate that even the recovery of signals with amplitudes of 0.3 ms$^{-1}$ with a period of 200 days is possible. Our tests also indicate that noise modelling can greatly improve the sensitivity of RV searches to low-amplitude signals. For instance, while the practise of nightly binning might remove correlations from the noise and make the excess noise appear ''whiter``, it also artificially decreases the amount of data and may lead to the undesirable side-effect that low-amplitude signals cannot be detected in the data after all.

We find that after removal of the moving average noise components in the HARPS data for HD10700, three signals can be discerned. Although, these signals can not be confirmed independently in the AAPS and Keck datasets, they are found confidently in the combined HARPS, AAPS, and HIRES data set as well. These three signals can also be found when using slightly modified (sub-optimal w.r.t. the preferred noise model of the HARPS data with MA(10) term and Gaussian white noise component) noise models, such as the MA(5) and MA(7). Also, two additional signals become statistically significant in the combined dataset. These signals discerned with Bayesian techniques and periodogram analysis are found at periods of 13.9, 35.4, 94, 168, and 630 days and do not have activity-induced counterparts based on the S and BIS indices from HARPS. The strongest of these signals at 35.4 days is close to the 34 day rotational period of the star which is a cause for concern. However, in addition to the lack of activity-induced signals near the rotation period, we also note that the period distribution in the posterior densities gives $\sigma \sim$ 0.0334 (see Fig. \ref{densities}) for the 35.362 day signal. If this signal was indeed induced by stellar rotation and activity, we would expect it to receive a much larger uncertainty under the reasonable assumption that a magnetic cycle and differential rotation were present. Any spot induced RV signal would be expected to change periodicity depending on its emergent latitude (which varies between 0 and 40 degrees on the Sun during an 11 year magnetic cycle), due to the latitude dependent differential rotation. We note that the HARPS data set has a baseline of 5.9 years (13.5 years for the combined data set), a significant proportion of the {\em Solar} activity cycle. The degree of rotational shear can be written as $\Omega(\theta) = \Omega_{eq} - \Delta\Omega \sin^{2}(\theta)$, where $\theta$ is the stellar surface latitude, $\Omega$ the angular velocity, and $\Omega_{eq}$ the equatorial angular velocity. Following \citet{barnes2005}, who found that $\Delta\Omega$ scales with stellar mass but is only weakly dependent on rotation rate, we obtain $\Delta\Omega = 0.06$ rad day$^{-1}$ or equivalently, $\Delta P$ = 11.04 days. This represents the maximum rotation period variation (i.e. between equator and polar regions), but for the solar-like case with spots limited to 0 and 40 degrees we expect to measure a rotation period variation of $11.04 \sin^{2}(40) = 4.56$ days. This would lead to a considerably larger $\sigma$ in the posterior density distribution than we find, even if we take into consideration that the HD10700 data only span the equivalent of approximately half a solar cycle where the maximum variation on latitude might be closer to 3.3 days. This is still an order of magnitude greater than the uncertainty in the obtained period. In other words, our posterior period distribution width for the 35.4 day signal is {\em not} consistent with an activity induced signal that a solar-like star might be expected to produce.

Similar arguments can be applied to the constrained nature of the other signals in the posterior densities (see Fig. \ref{densities}) and thus we do not consider that rotation and spots are a ready explanation for the signals. Thus it is plausible that the signals are caused by a system of five planets with minimum masses of 2.0, 3.1, 3.6, 4.3, and 6.6 M$_{\oplus}$, respectively. Indeed, such a system would be dynamically stable. However, before going farther with a planetary explanation, it should be emphasised that (1) the periodic signals that we are discussing appear following the removal of the ''moving average`` noise apparent in the data and with the assumption of Gaussian errors and that (2) with the available data we cannot discover any of the signals independently in more than one different dataset. It is worth noting that all the signals have similar MAP amplitudes between 0.58 and 0.75 ms$^{-1}$. Given the nature of these signals appearing after removal of noise it is unsurprising that they all have low amplitudes. However, for a planetary interpretation it might require some coincidence that the planets in this system have such an arrangement that the masses and periods conspire to correspond to similar Doppler amplitudes. Although, our simulations indicate that the recovery of such low-amplitude signals is not particularly dependent on the details of the moving average noise removal, the actual recovered amplitude of such small signals in our simulations might be overestimated. Thus, care must be taken in the interpretation of the nature of the signals.

We note that there are many possible ways to model the noise in radial velocity data and that in this work we have only explored a few of them. Indeed, the signals we detect may also result from the combination of insufficient noise modelling and our lack of understanding of stellar physics: asteroseismology, starspots, magnetic cycles, granulation and other phenomena of stellar surfaces.

In the future, we hope to expand the quantity of high precision radial velocity data of HD 10700 and to consider a wider range of noise models and incorporate an improved understanding of stellar phenomena. While our noise model choices serve to account for asteroseismological noise on timescales at or near the exposure times they do not necessarily correspond physically to e.g. granulation noise in the data. Comparisons of a more extensive collection of statistical models taking into account non-white features in the RV noise are needed to verify or falsify the existence of the signals we report in this work. Also, future data will be of essence in determining the nature of the signals we detect. If the five periodicities can be detected independently in different datasets, their genuine nature as signals of stellar origin will be verified. While even this will not imply that they are definitely Doppler signatures of low-mass planets, it will help ruling out spurious periodicities that insufficient modelling and instrumental instability might cause.

With a distance of only 3.7 pc, HD 10700 is the third closest star reported to be a host to a putative planetary system after $\epsilon$ Eridani \citep{hatzes2000} with a distance of 3.2 pc and $\alpha$ Centauri B \citep{dumusque2012} with a distance of 1.3 pc, though both of these remain to be confirmed and \citet{zechmeister2012} have casted considerable doubt on the existence of a planet around $\epsilon$ Eridani. This makes HD 10700 an ideal target for future direct-imaging missions. The signals we find, which suggests the presence of low-mass planets, are consistent with both current theoretical models for low-mass planet formation and extant observational evidence for the presence of low-mass planets in the immediate Solar neighbourhood. Assuming the signals are indeed of planetary origin, the orbital periods of the two innermost candidates appear to suggests a 5:2 commensurability that would likely enable the long-term stability of the system \citep[see e.g. a similar stabilising resonance of NN Serpentis][]{horner2012}. Given this assumption, we also note that a candidate corresponding to the signal with a period of 168 days would have an orbit inside the liquid water habitable zone as defined by \citep{selsis2007}. However, these issues remain merely speculative until the planetary origin of the signals can be verified by an independent detection.

\begin{acknowledgements}
M. Tuomi is supported by RoPACS (Rocky Planets Around Cool Stars), a Marie Curie Initial Training Network funded by the European Commission's Seventh Framework Programme. J. S. Jenkins acknowledges funding by Fondecyt through grant 3110004 and partial support from Centro de Astrof\'isica FONDAP 15010003, the GEMINI-CONICYT FUND and from the Comit\'e Mixto ESO-GOBIERNO DE CHILE. S. Vogt gratefully acknowledges support from NSF grant AST-0908870. J. Horner and C. G. Tinney gratefully acknowledge the financial support of the Australian government through ARC Grant DP0774000. The authors acknowledge the significant efforts of the HARPS-ESO team in improving the instrument and its data reduction pipelines that made this work possible. We also acknowledge the efforts of all the individuals that have been involved in observing the target star with HARPS, Keck, and the AAT. The work herein is based in part on observations obtained at the W. M. Keck Observatory, which is operated jointly by the University of California and the California Institute of Technology. Finally, the authors would like to thank the anonymous referee for comments and suggestions that enabled considerable improvements of the manuscript.
\end{acknowledgements}

\appendix

\section{Phase-folded orbits}

\begin{figure*}
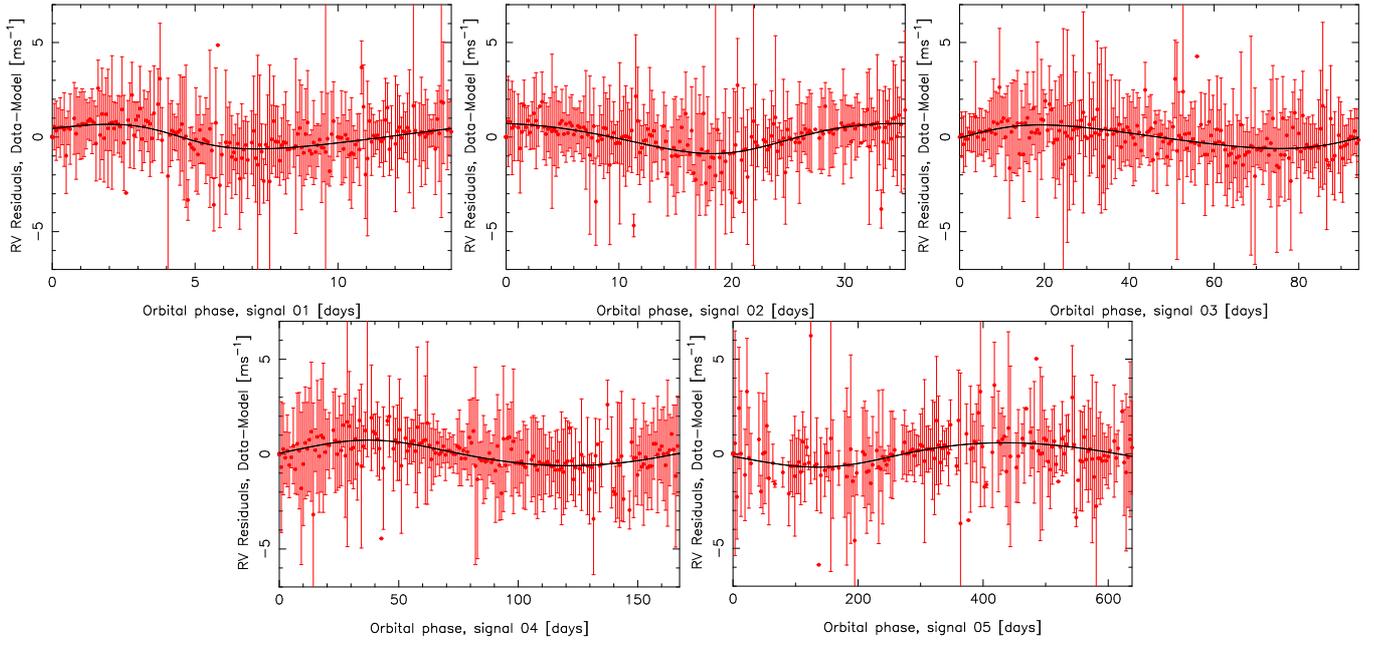

\center
\includegraphics[angle=-90,width=0.32\textwidth]{rvdist05_scresidd_rv_HD10700_1.ps}
\includegraphics[angle=-90,width=0.32\textwidth]{rvdist05_scresidd_rv_HD10700_2.ps}
\includegraphics[angle=-90,width=0.32\textwidth]{rvdist05_scresidd_rv_HD10700_3.ps}

\includegraphics[angle=-90,width=0.32\textwidth]{rvdist05_scresidd_rv_HD10700_4.ps}
\includegraphics[angle=-90,width=0.32\textwidth]{rvdist05_scresidd_rv_HD10700_5.ps}
\caption{Phase-folded Keplerian signals for the combined HARPS, HIRES, and AAPS data. Orbital phase of each signal is divided into 200 bins and the means and the corresponding standard deviations of the data in each bin are plotted together with the Keplerian signal.}\label{bin_orbits_all}
\end{figure*}

\begin{figure*}
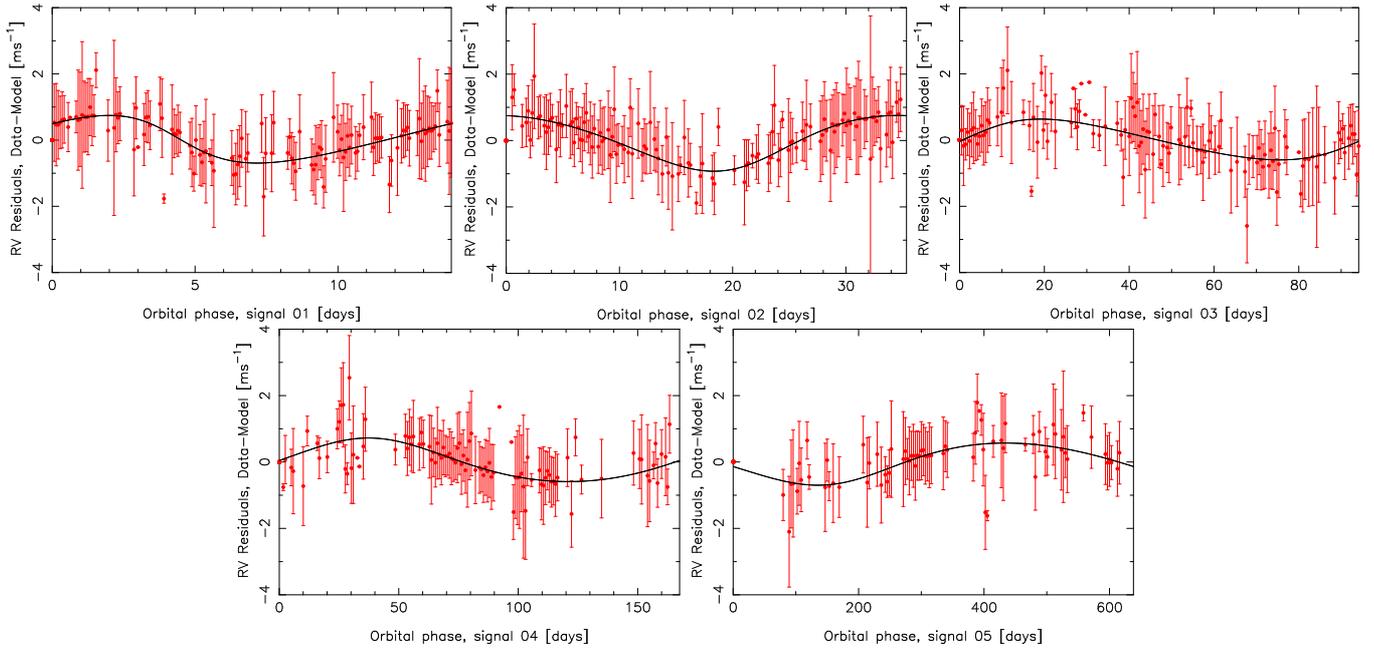

\center
\includegraphics[angle=-90,width=0.32\textwidth]{rv_HARPSdist05_scresidd_rv_HD10700_1.ps}
\includegraphics[angle=-90,width=0.32\textwidth]{rv_HARPSdist05_scresidd_rv_HD10700_2.ps}
\includegraphics[angle=-90,width=0.32\textwidth]{rv_HARPSdist05_scresidd_rv_HD10700_3.ps}

\includegraphics[angle=-90,width=0.32\textwidth]{rv_HARPSdist05_scresidd_rv_HD10700_4.ps}
\includegraphics[angle=-90,width=0.32\textwidth]{rv_HARPSdist05_scresidd_rv_HD10700_5.ps}
\caption{As in Fig. \ref{bin_orbits_all} but for the high-precision HARPS data alone.}\label{bin_orbits_harps}
\end{figure*}


\section{Radial velocities}

\begin{table}
\center
\caption{AAPS radial velocities of HD 10700.}\label{tab:rv_aaps}
\begin{tabular}{lcc}
\hline \hline
Time & Velocity & Unc. \\

[JD-2440000] & [ms$^{-1}$] & [ms$^{-1}$] \\
\hline
10830.993 & -5.09656 & 2.17488 \\
10830.994 & -12.0951 & 2.27151 \\
11118.076 & -0.0715645 & 2.17066 \\
11121.078 & 3.23342 & 2.23103 \\
11212.938 & -2.13859 & 1.93977 \\
11235.904 & -8.47236 & 3.14823 \\
11235.905 & -3.59754 & 1.95095 \\
11236.904 & -0.486371 & 2.3019 \\
11383.336 & -1.02518 & 1.42745 \\
11383.338 & -6.70739 & 1.4056 \\
11385.34 & -2.12584 & 3.70331 \\
11385.342 & -1.56196 & 3.25338 \\
11386.336 & 1.01967 & 1.87225 \\
11386.337 & -2.38868 & 1.55173 \\
11387.338 & -3.91093 & 1.37435 \\
11413.162 & -6.1967 & 1.67232 \\
11414.319 & -0.525862 & 1.55716 \\
11473.109 & -3.10539 & 1.86229 \\
11526.966 & -0.500644 & 1.90503 \\
11527.93 & 0.48339 & 1.90033 \\
11743.343 & 0.557373 & 1.45827 \\
11767.286 & 0.408715 & 2.08542 \\
11767.288 & -9.98603 & 2.23851 \\
11767.289 & 0.0377612 & 1.9731 \\
11768.29 & -5.93756 & 2.15648 \\
11768.292 & -6.00748 & 1.78525 \\
11828.097 & -3.31695 & 2.0453 \\
11829.009 & -8.66755 & 2.01648 \\
11829.985 & -4.84257 & 2.10931 \\
11856.095 & -4.72711 & 2.05964 \\
11856.928 & -24.1846 & 5.58004 \\
11918.986 & -4.82101 & 2.17249 \\
11919.978 & -5.08535 & 1.81794 \\
12061.344 & -5.65156 & 2.41785 \\
12092.276 & 0.598991 & 1.49112 \\
12093.356 & 1.23776 & 1.56277 \\
12093.358 & -2.02704 & 1.55199 \\
12093.359 & 1.9919 & 2.03245 \\
12093.36 & -7.49353 & 2.14144 \\
12125.23 & -9.08186 & 1.54394 \\
12125.233 & -4.46409 & 1.39802 \\
12125.236 & -3.97475 & 2.73834 \\
12125.239 & -4.11543 & 2.06392 \\
12125.242 & 3.28086 & 1.934 \\
12125.245 & -8.3693 & 2.22455 \\
12125.248 & -1.78719 & 1.51135 \\
12126.111 & -5.68895 & 1.62347 \\
12126.113 & -4.76531 & 1.46686 \\
12126.115 & -2.23099 & 1.47897 \\
12126.117 & -1.02592 & 1.3687 \\
12126.119 & -0.874087 & 1.46358 \\
12128.35 & -2.95253 & 2.15584 \\
12128.352 & -7.8229 & 2.1805 \\
12128.353 & -3.99854 & 2.32461 \\
12128.355 & -11.0018 & 1.98795 \\
12130.351 & -5.32129 & 1.88677 \\
12130.352 & -8.89184 & 1.77843 \\
12130.354 & -10.2658 & 2.25454 \\
12151.321 & -8.25988 & 3.69202 \\
12151.322 & 2.55054 & 3.26951 \\
12151.323 & -7.98841 & 2.1308 \\
12151.324 & -1.08759 & 2.26264 \\
12190.185 & -0.757956 & 2.03916 \\
12190.189 & -7.50007 & 2.6145 \\
12190.193 & -3.79168 & 4.02458 \\
12190.197 & 0.878117 & 2.33586 \\
12421.339 & -6.98129 & 1.48632 \\
12421.342 & -6.36269 & 1.47724 \\
12422.337 & -7.72013 & 1.76449 \\
12422.339 & -3.54478 & 1.57046 \\
12423.338 & -5.73116 & 1.60975 \\
12423.339 & -5.61929 & 1.51408 \\
12424.343 & -4.82933 & 1.73107 \\
12424.344 & -5.94701 & 1.57142 \\
12425.34 & -5.55552 & 1.53549 \\
12425.341 & -0.936324 & 1.55162 \\
12454.347 & 0.915797 & 1.64056 \\
12454.349 & -4.94121 & 1.60846 \\
12454.35 & -0.938487 & 1.99133 \\
12455.349 & 3.17405 & 1.35904 \\
12455.35 & 2.65388 & 1.36533 \\
12456.344 & 2.63021 & 1.61626 \\
12477.313 & -0.410969 & 1.65665 \\
12510.278 & -2.92027 & 1.69886 \\
12510.282 & -1.61716 & 1.50267 \\
12511.139 & -1.48691 & 1.65459 \\
12511.141 & -4.67084 & 1.73124 \\
12511.142 & -3.15409 & 1.7342 \\
12511.144 & -1.23425 & 1.68165 \\
12591.997 & -2.58149 & 1.86751 \\
12591.998 & 2.52396 & 1.7552 \\
12591.999 & 2.69175 & 1.90252 \\
12593.038 & 2.51661 & 1.74913 \\
12593.042 & 0.296852 & 1.74951 \\
12593.046 & 1.177 & 1.66736 \\
12593.052 & 2.02497 & 1.72143 \\
12593.058 & 1.86904 & 1.62212 \\
12593.063 & 2.85686 & 1.84517 \\
12593.067 & -1.18688 & 1.99375 \\
12593.069 & -1.78776 & 2.46208 \\
12593.07 & -3.01894 & 5.183 \\
12593.081 & -2.17612 & 1.84035 \\
12593.087 & 3.15944 & 3.33451 \\
12594.027 & 1.25385 & 1.81738 \\
12594.029 & -0.588582 & 1.89613 \\
12594.031 & 0.102355 & 1.81155 \\
12595.023 & -2.88425 & 1.96782 \\
12595.024 & -2.18509 & 1.87146 \\
12595.026 & -2.97544 & 2.10707 \\
12598.085 & 0.902761 & 1.67372 \\
12598.087 & -0.782177 & 1.59916 \\
12598.088 & -1.96997 & 1.7646 \\
12599.046 & -3.42237 & 1.85671 \\
12599.05 & -2.24683 & 2.5893 \\
12599.054 & -0.949761 & 2.26776 \\
12599.067 & -10.3627 & 3.05592 \\
12599.071 & -2.97971 & 1.97037 \\
12599.077 & -0.29657 & 1.65944 \\
12599.083 & -1.72797 & 1.82065 \\
12599.088 & 1.24048 & 1.58828 \\
12600.113 & 2.92653 & 1.83158 \\
12600.118 & 2.22013 & 1.68245 \\
12600.122 & -0.137515 & 1.66128 \\
12654.909 & 3.14385 & 2.66874 \\
12654.911 & -2.70792 & 1.63968 \\
12708.885 & 21.1535 & 5.26155 \\
12708.887 & 0.229482 & 5.10769 \\
12708.889 & 3.23328 & 5.29917 \\
12710.876 & -2.37555 & 1.69784 \\
12710.877 & 0.167767 & 1.72142 \\
12710.878 & -2.56719 & 1.74207 \\
12711.883 & 3.8403 & 1.51644 \\
12711.884 & 2.45352 & 1.53227 \\
12711.886 & 3.96421 & 1.49485 \\
12784.336 & -2.05076 & 1.68306 \\
12784.337 & 0.29448 & 1.61529 \\
12784.338 & -2.06328 & 1.7151 \\
12784.339 & 1.55493 & 1.63014 \\
12785.336 & -6.845 & 8.06261 \\
12785.337 & -9.53695 & 3.50682 \\
12785.339 & -1.48853 & 2.71783 \\
12857.287 & -12.3489 & 2.03114 \\
12857.289 & 0.687823 & 2.12462 \\
12857.29 & -3.23451 & 1.78123 \\
12857.292 & -4.15573 & 1.71759 \\
12858.195 & -19.0414 & 5.44323 \\
12858.197 & -1.19626 & 3.5238 \\
12858.199 & -2.59552 & 5.54601 \\
12858.201 & -2.50853 & 2.63455 \\
12858.203 & -3.5776 & 2.13693 \\
12858.25 & -1.30485 & 2.22545 \\
12858.252 & -1.60604 & 2.53154 \\
12858.255 & -1.93322 & 2.19491 \\
12858.257 & -2.97973 & 2.25681 \\
12859.26 & -3.22037 & 1.57997 \\
12859.262 & -2.70588 & 1.41838 \\
12860.34 & 3.32827 & 1.5079 \\
12860.342 & 1.92628 & 1.44367 \\
12860.343 & -0.216214 & 1.58404 \\
12860.344 & -1.99781 & 1.50237 \\
12860.346 & 0.651604 & 1.55807 \\
12861.341 & 0.123206 & 1.53235 \\
12861.343 & 4.54828 & 1.5931 \\
12943.119 & 1.79781 & 1.71286 \\
12943.12 & -0.703808 & 1.80386 \\
12943.982 & 2.521 & 3.27066 \\
12943.985 & 0.388788 & 4.31655 \\
12945.131 & 4.12968 & 2.07946 \\
12945.132 & 3.55521 & 1.91061 \\
12946.077 & 5.7226 & 2.33338 \\
12946.078 & 14.7287 & 3.32391 \\
12947.073 & 20.8155 & 2.80612 \\
12947.074 & 9.57653 & 1.98654 \\
13003.959 & -3.17731 & 1.7922 \\
13003.96 & -0.626619 & 1.91029 \\
13007.057 & -0.937858 & 1.85723 \\
13007.058 & 0.426439 & 1.83999 \\
13007.991 & -7.15131 & 1.92793 \\
\hline \hline
\end{tabular}
\end{table}

\end{document}